\documentclass[useAMS,usenatbib]{mn2e}
\usepackage{times}
\usepackage{graphicx}
\usepackage{amssymb}
\usepackage{natbib}
\usepackage{psfig}
\title[The Globular clusters-stellar haloes connection in early type galaxies]{The globular clusters-stellar haloes connection in early type galaxies}
\author[Forte et al.]{Juan C. Forte$^{1,2}$\thanks{E-mail:forte@fcaglp.unlp.edu.ar}, E. Irene Vega$^{1,2}$ and Favio Faifer$^{1,2,3}$ \\
$^1${Facultad de Ciencias Astron\'omicas y Geof\'isicas, Universidad Nacional de La Plata}\\
$^2${Consejo Nacional de Investigaciones Cient\'ificas y T\'ecnicas, Rep.
Argentina}\\
$^3${IALP}\\
}
\begin{document}

\date{Accepted . Received ; in original form }
\pagerange{\pageref{firstpage}--\pageref{lastpage}} \pubyear{}
\maketitle

\label{firstpage}

\begin{abstract}

   This paper explores if, and to what an extent, the stellar populations of 
   early type galaxies can be traced through the colour distribution of
   their globular cluster systems. The analysis, based on a galaxy sample from the
   Virgo ACS data, is an extension of a 
   previous approach that has been successful in the cases of the giant
   ellipticals NGC 1399 and NGC 4486, and assumes that
   the two dominant GC populations form along diffuse stellar populations
   sharing the cluster chemical abundances and spatial distributions.
   The results show that a) Integrated galaxy colours can be matched to
   within the photometric uncertainties and are consistent with a narrow
   range of ages; b) The inferred mass to luminosity ratios and stellar
   masses are within the range of values available in the
   literature; c) Most globular cluster systems occupy a thick plane in
   the volume space defined by the cluster formation efficiency, total
   stellar mass and projected surface mass density. The formation efficiency
   parameter of the red clusters shows a dependency with projected stellar
   mass density that is absent for the blue globulars. In turn, the
   brightest galaxies appear clearly detached from that plane as a possible
   consequence of major past mergers; d) The stellar mass-metallicity relation is
   relatively shallow but shows a slope change at $M_*\approx 10^{10} M_\odot$.
   Galaxies with smaller stellar masses show predominantly unimodal globular
   cluster colour distributions. This result may indicate that less massive
   galaxies are not able to retain chemically enriched intestellar matter.

\end{abstract}

\section{Introduction}
\label{INTRO}

    Pioneering ideas about the role of globular clusters (GC) as tracers
    of the formation history of galaxies come from \citealt{EGG62}
    or, with a different perspective, from  \citealt{SZ78}. The
    evolution of this subject during recent years has been reviewed in \citet{BS06}.
    A key issue in that context is the connection between GC and the ``diffuse'' stellar
    populations of the galaxies they are associated with. Some reasons allow
    to infer that such a link might not be simple. 
    However, several papers point out that massive early
    type galaxies  seem to have highly synchronised ages. For example,
   \citet{LBA08} reach
    that conclusion in their discussion of the properties of the Fundamental Plane
    for early type galaxies also indicating that the age variation per mass decade 
    seems smaller than a few percent.
    This suggests that the eventual connection between field stars and
    GC may still be ``read'' since the effects of subsequent (younger) stellar
    populations in these galaxies may have not been important.

    The theoretical side of the problem has received a number of
    contributions in an effort to better define its cosmological
    background (e.g.  \citealt{BEA02} or,more recently, \citealt {BEK08}).
    Differences and similarities between GC and the underlying stellar
    populations have been pointed out in the past. For example, Forte,
    Strom and Strom (1981) found that, in the average, GC appear
    significantly bluer than the stellar halos at all galactocentric 
    radii in four Virgo ellipticals.

    In turn, and after the discovery of GC colour bimodality (e.g. \citealt{ZA93};
    \citealt{OGF93}) \citealt{FF01} noted that
    the stellar halo colours in the {\bf inner regions} of early type galaxies
    are remarkably similar to those of the ``red'' GC population.

    A more recent comparison between the integrated features of the NGC
    3923 galaxy halo and its GC population has been presented in \citealt{NO08}     
    on the basis of Gemini-GMOS spectra and direct imaging data. These
    authors show that the galaxy halo has a chemical abundance close to the
    average value defined by its GC and that both systems are coeval.

    A point to be considered in that kind of comparisons, however, is
    that the GC colour statistics are ``number weighted''  while
    halo colours are, naturally, ``luminosity'' weighted. Thus, a coincidence
    between GC and stellar halo colours would be expected only if each GC is 
    formed along a given  diffuse stellar mass on a constant luminosity basis.

    An alternative approach has been presented in \citealt{FFG05}
    and \citealt{FFG07} (hereafter FFG07)
    in a comparative study of the giant ellipticals
    NGC 1399 and NGC 4486 (M87). The main assumption in these works is
    that GC are good tracers of the spatial distribution and chemical
    abundance spread of ``diffuse'' stellar populations, formed along with
    the clusters, and that the number of GC,
    $N$, per diffuse stellar mass unit $M_*$ with  chemical abundance $Z$,
    is given by

\begin{equation}
dN(Z)/dM_*(Z) = {\gamma} \exp({-{\delta  [Z/H])}}\
\end{equation}

    \noindent which is a generalization of the \citet{ZA93} {\bf T} parameter. In 
    those galaxies, the results show that the number of GC per diffuse
    stellar mass increases when metallicity decreases.

    $\gamma$ and $\delta$ are free parameters that, once the spatial distribution
    of the GC and their chemical abundance scales are determined, allow a
    fit of the galaxy surface brightness profiles over several tens of Kpc. These
    fits also provide a good approximation to the galactocentric colour gradients
    as well as of the difference between galaxy halo colours and GC mean colours.
    In turn, the {\bf inferred} metallicities for field stars show a remarkable similarity
    with the observed ones in nearby resolved galaxies.

\begin{figure*}
\resizebox{0.5\hsize}{!}{\includegraphics{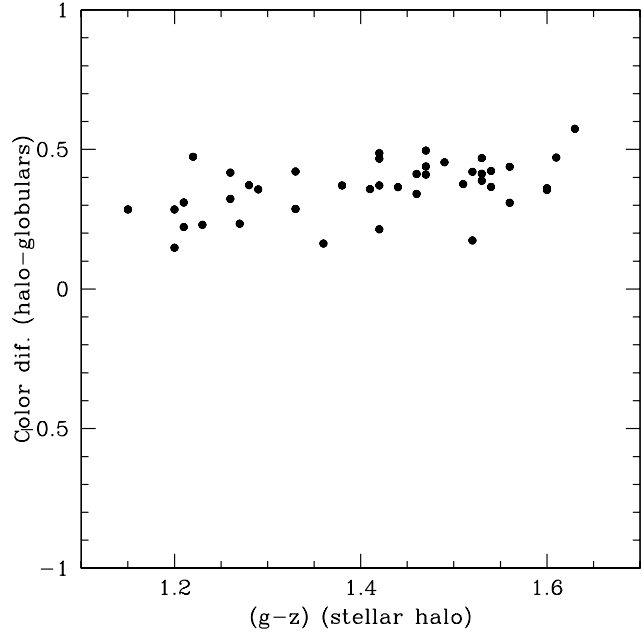}}
\caption{ 
 $(g-z)$ colour difference between galaxy halo colours (from Ferrarese
   et al. 2006) and the average GC colours in each galaxy (from Peng et
   al. 2006).}  
\label{halo_globs.dif}
\end{figure*}

    This paper aims at further testing the modeling described above,
    taking advantage of the largest and most homogeneous GC data set
    available so far in the literature: The Virgo Advanced Camera Survey
    (\citealt{Cote04} and subsequent papers). In particular, we refer to
    the galaxy photometry presented by \citealt{FE06}; hereafter F2006) and
    to the GC systems study by \citealt{Peng06}; hereafter P2006
    ). Figure \ref{halo_globs.dif} shows the colour difference between stellar halo
    colours and mean GC colours for a galaxy sub-sample (see paragraph \ref{HISTOFITS}) 
    from F2006 and P2006. This diagram is consistent with the results in 
    Forte et al. (1981) mentioned above.
%
 
    A thorough discussion about the characteristics of the GC formation
    efficiencies in terms of  galaxy brightness, mass, and environment within the
    Virgo cluster has been recently presented by  \citealt{Peng08}. This work
    shows that both the number of GC per galaxy brightness or galaxy mass units
    have a non monotonic behaviour
    reaching a minimum for galaxies with stellar masses $\approx 5 . 10^{10} M_\odot$ .

    That overall trend, in turn, is compatible with the results obtained by \citet{RZS05}
    and later by 
\citet{SPI}, in their discussion of galaxies more massive than that mass, and also with
those by \citet{ML} in the analysis of GC systems associated to less massive dwarf galaxies.

    In this paper we address the same subject although with some differences 
    in comparison with \citealt{Peng08}, namely:

    -GC colour histograms are decomposed in terms of ``blue'' and ``red''
    clusters using the FFG07 approach instead of the
    non-parametric fits performed by P2006.

    -Stellar mass to B luminosity ratios are derived on the basis of the
    FFG07 modeling and used to estimate stellar masses within a
    galactocentric radius of $\approx$ 150 arcsecs, the typical coverage of the
    Virgo ACS.

    -Cluster formation is analyzed in the volume space defined by GC
    formation efficiency, galaxy stellar mass, and projected stellar mass
    densities, as well as in terms of  chemical abundances.

 \section{Model assumptions (and caveats)}
 \label{MOD}

 The GC-stellar haloes link given in FFG07 involves several assumptions,
 namely:

 a) GC colour bimodality in fact reflects two different (``blue'' and ``red'')
 GC  families which have distinct chemical abundance and spatial
 distributions. Several arguments support this assumption (see \citealt{KZ07};
 \citealt{SBB}).

  b) GC colour histograms are usually decomposed via two-Gaussian fits
  that involve five free parameters. Instead, FFG07 seek for a
  link betweeen chemical abundance $Z$ and integrated colour by means of
  an empirical colour-metallicity relation (see paragraph \ref{COLMET}).
  Exploring different trial functions leads to $N(Z) \approx  exp(Z-Zi)/Zs$,
  where $Zi$ is an initial chemical abundance, as the simplest approximation
  for the colour histogram decomposition. In this work, as in FFG07, we set
  $Zi=0.0035 Z_{\odot}$.
  The histogram fit then requires three parameters: the chemical
  abundance scale lenghts  $Zs(blue)$, $Zs(red)$ and the
  relative fraction of clusters belonging to each GC family.

  The case of the blue GC deserves a particular comment in relation with
  the presence of the so called ``blue tilt'', i.e., these clusters seem to
  become redder with increasing brightness in some galaxies. This effect 
  produces a spread of the blue GC colours that FFG07 identify as a 
  variation of $Zi$ with cluster mass in the case of NGC 4486.

  We note that the existence of the tilt in NGC 4486, is a 
  subject of debate in the literature, e.g., \citet{St06}; \citet {Mie06}, 
  FFG07, \citealt{KU08}, or \citealt{WZ08}.

 c) Blue and red GC families are assumed to be old and formed within a relatively
  narrow period of time. Arguments supporting this assumption can be found, for
  example, in \citet{BS06} (see their figure 7).

  Alternatively, some intermediate age globular candidates have been
  reported, among the ``red'' globulars in other galaxies, by \citet{Forb01} (also
  see \citealt{Pi06a}, \citealt{Pi06b}; \citealt{He07}).

 d) The ratio of the number of GC to diffuse stellar mass with the same
  $Z$, is assumed to follow the exponential relation given in paragraph
  \ref{INTRO}. For typical values of $\bf \delta$, that ratio increases when chemical
  abundance decreases. This is a ``phenomenological'' assumption, that
  allows a good fit to the surface brightness and halo colour gradients
  in NGC 1399 and NGC 4486, and that we confront with the Virgo ACS data in
  this work.

  A caveat in the whole approach concerns to the fact that GC statistics
  refer to clusters that have survived to a number of destructive events of
  dynamical origin (\citealt{CL95}; \citealt{GO97}), i.e.,
  these GC might not reflect the original cluster populations. For example,
  both the blue and red GC show cores in their projected galactocentric surface
  densities  (although with widely different spatial scales) that may be the result
  of that depletion process. These cores contrast with the peaked brightness
  profiles in the inner region of galaxies.

\section{The colour metalicity calibration }
\label{COLMET}

   An empirical $(C-T_1)$ colour versus $[Fe/H]_{zw}$ metallicity relation (on the
  \citealt{ZW84} scale) was presented in FFG07:

\begin{equation}
(C - T_1) = 0.94 + 0.068~([Fe/H]_{zw} + 3.5)^2 
\end{equation}

\begin{figure}
\resizebox{1.0\hsize}{!}{\includegraphics{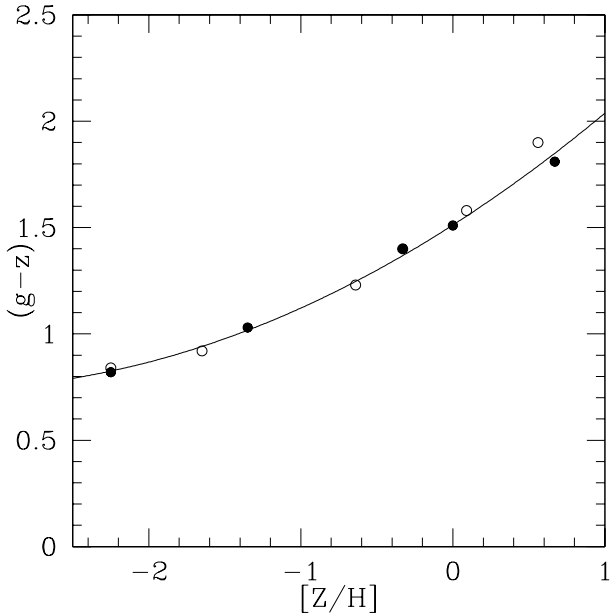}}
\caption{
 Integrated $(g-z)$ colour versus chemical abundance from SSP models for an
  age of 12 Gy. Filled circles: \citet{MA04}); open circles: 
 \citet{BC03}. The continuous line is the empirical calibration
  given by FFG07, adopting $[Z/H]=[Fe/H]_{zw}+0.131$, and 
  $(g-z)=(C-T_1)-0.20$ (see text).}

\label{colormet}
\end{figure}

  This relation was obtained by combining 100 GC in the MW and
  98 in other three galaxies whose metallicities were derived
  from Lick indices. This relation is also adopted in this work
  after transforming the $(g-z)$ colours to $(C-T_1)$ as discussed in
  paragraph \ref{PHOTSC}. A comparison with SSP models  by \citet{MA04} 
  and \citet{BC03}, for an 
  age of 12 Gy, is shown in Figure \ref{colormet}.

 Considering the $(g-z)$ and $(B-I)$ colours in the \citet{MA04} models, that leads to:

\begin{equation}
(g - z) = 1.00~(B - I) - 0.68
\end{equation}

\noindent and the  empirical $(B-I)$ versus $(C-T_1)$ colours relation for MW clusters
  given in FFG07:

\begin{equation}
(C - T_1)  = 1.03~(B - I) - 0.43
\end{equation}

\noindent indicates that $(C - T_1)$ and $(g - z)$ can be linked just by means of
  a zero point difference, a procedure that we describe and adopt in the following
  paragraphs.

\section{Photometric scales}
\label{PHOTSC}

  For the sake of homogeneity with FFG07, we first transform the
  $g$ magnitudes given by F2006 to $B$ magnitudes (Johnson system).
  The $B$ luminosities, in turn, will be used to estimate the galaxy
  masses though the $(M/L)_{B}$ ratios derived from fits to the GC 
  colour histograms.

  The Virgo ACS survey has a field coverage of 3.4 arcmin,
  which roughly means a galactocentric range of $\approx 150$ arcsecs. Galaxy
  $g$ magnitudes were then obtained within an elliptical contour, with
  a semimajor axis {\bf a}= 150 arcsecs, and
  using the S\'ersic best fit parameters given  in F2006
  (table 3). In turn, integrated $B$ magnitudes, within that galactocentric range,
   were obtained for 14 galaxies, fainter than $B$=11.0 mag, in common with \citet{CAON90}
   yielding:

\begin{equation}
B = g(150) + 0.28~(\pm 0.19/\sqrt{14})
\end{equation}

\noindent a transformation that we adopt in what follows.

  This connection between the $g(150)$ and $B$ magnitudes can be also
  compared with that derived from the relations given by \citet{ROD06} (based on stellar
  photometry), and adopting a mean colour $(g-z)$=1.3 for the ACS galaxies, leading to
  $B= g(150)+0.44$. The differences between both zero points give an idea about the
  photometric uncertainties involved in this kind of analysis. 

\begin{figure}
\resizebox{1.0\hsize}{!}{\includegraphics{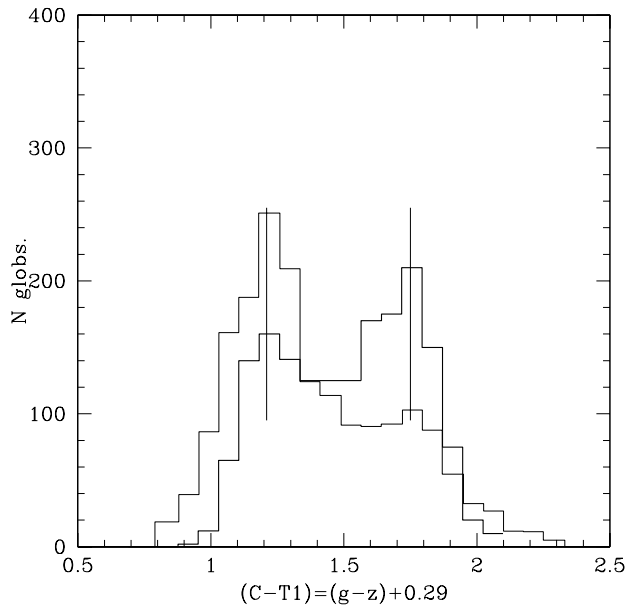}}
\caption{
$(C-T_1)$ colour histogram for the NGC 4486 globular cluster candidates
  with galactocentric distances from 120 to 360 arcsecs (thin line) using
  the data from FFG07. The thick line corresponds to clusters within
  150 arcsecs from the galaxy center as given by Peng et al. 2006. The
  binning steps are those defined in this last work. Vertical lines
  show the positions of the ``blue'' and ``red'' peaks. $(g-z)$ was 
  transformed to $(C-T_1)$ by adding 0.29 mags.(see text).}

\label{comphisto}
\end{figure}

  The FFG07 modelling delivers $(C-T_1)$ colours that can be transformed
  to $(g-z)$ just by adopting a colour offset (as discussed in the
  previous paragraph). First we look for the link between the
  $(C-T_1)$ and $(g-z)$ colour scales from FFG07 and P2006, respectively,
  and using the photometry of the well populated GC system of NGC 4486 
  (VCC 1316).
  Figure \ref{comphisto}
  shows the $(C-T_1)$ colour  histogram for GC within a galactocentric range of 120
  to 360 arcsecs from FFG07. This diagram also shows the GC histogram given
  by P2006 for the inner regions of the same galaxy, and adopting a shift

\begin{equation}
(C - T_1) - (g-z)~(P2006) = 0.29
\end{equation}

\noindent that agrees with a previous estimate in FFG07. Vertical lines in Figure 3
  indicate that the position of the ``blue'' and ``red'' GC peaks are consistent
  in both statistics (using the same binning steps as in P2006) and no variation
  of these peaks seem detectable with galactocentric radius. The same feature
  holds further out  in galactocentric radius as shown by \citet{KZ07}. We note
  that the blue tail of the GC colour distribution is more populated in the outer
  field, suggesting a somewhat bluer lower boundary in colour. The significance of
  this effect is hard to asses since the outer region is more likely to be affected by
  field interlopers (for example, unresolved blue galaxies).

      
  A similar approach was adopted regarding the galaxy integrated
  colours. F2006 give $(g-z)$ = 1.60 for NGC 4486 while the integration of the
  $B$ profile from Caon et al. (1990), combined  with the (B-R) colour gradient
  listed by Michard (2000), leads  to $(B-R)$ = 1.53 within a galactocentric radius
  of 150 arcsecs. Given the GC colour relations in FFG07, this colour
  transforms to ($C - T_1)$ = 1.77 and then:
\begin{equation}
(C - T_1) - (g - z)~(F2006) = 0.17
\end{equation}

  The zero point difference between the F2006 and P2006 colour scales is coherent
  with the results given by \citet{JALI09} in their analysis of the (g-z) colour
  magnitude relation of early type galaxies.

\section{GC Colour Histogram fits}
\label{HISTOFITS}

  P2006 present $(g-z)$ colour histograms for GC systems
  in 100  Virgo galaxies (their figure 2). Our analysis is restricted
  to galaxies with more than $\approx$ 15 GC and at least five
  colour bins (after removing background contamination). Smaller cluster populations, or
  lower definition colour histograms, prevent  meaningful fits.

  There are four galaxy fields (P2006 numbers: 43, 61, 79, 93) that, 
  being close to galaxies with large CG populations, might be likely
  contaminated by their GC. In turn, galaxy number 39 is the only unimodal galaxy
  with a  narrow red peak. This last GC system, and those in
  galaxies 27, 32, 33 and 65, could not be properly fit due to noisy
  histograms or high background contamination. As a result of the
  criteria mentioned above, and after rejecting these last objects,
  our sample finally comprises 63 galaxies listed in Table\,\ref{sample}.   

  P2006 perform a non-parametric fit to the GC
  histogram which, as mentioned, is different from the approach
  presented in this paper. Here we adopt a Monte Carlo based approach
  that seeks the determination of the chemical abundance scales, $Zs(blue)$ and
  $Zs(red)$, as well as the fraction of clusters in each GC population
  from the GC colour histograms.

  First, a ``seed'' globular is randomly generated with a given $Z$
  (controlled by an initial  $Zs$ parameter). This value is
  transformed to $[Fe/H]_{zw}$ and to  $(g-z)$ colour, by means
  of our colour-metallicity relation. Interstellar colour excesses
  and gaussian observational errors (with a dispersion $\pm$ 0.05 in $(g-z)$)
  are added to each cluster. The resulting synthetic histograms are then compared
  with those given by P2006 using the same variable binning steps
  adopted by these authors.

  The $Zs$ values, and the ratio of clusters belonging to the blue
  or red populations, are iterated until a ``quality fit'' $\chi^{2}$ parameter,
  as defined by \citet{Cote98}, is minimised.

  Throughout this analysis we use an average E(B-V) = 0.028, from
  the colour excesses  given by F2006 which, in turn, are based on the
  Schlegel et al. (1998) maps. This seems justified by the uncertainties
  of these maps although, as noted by \citealt {Peng08}, some noise would be
  expected on the derived blue luminosities arising in dust eventualy
  associated with each galaxy.

  The resulting $Zs$ scales, number of clusters assigned to each GC family and
  the quality of the fit indicator, are given in Table\,\ref{sample}. Typical
  uncertainties connected with counting statistics range from 0.005 to 0.01
  for $Zs(blue)$ and 0.05 to 0.15 for $Zs(red)$. 

\begin{table*}
\caption{Globular cluster colour histogram fit parameters for Virgo ACS galaxies. Col. [1]: Galaxy number from P2006; Col. [2]: VCC galaxy number from \citealt{BST85}; Col. [3]: Number of blue GC; Col. [4]:  Chemical abundance scale for the blue GC in $Z_\odot$ units; Col. [5]: Number of red GC; Col. [6]: Chemical abundance scale for the red GC in $Z_\odot$ units; Col. [7]: GC colour histogram quality fit parameter (as in \citealt{Cote98}); Col. [8]: Fraction of stellar mass associated with the red GC; Col. [9]: Integrated apparent g magnitude (within a=150 arcsecs); Col. [10]: Integrated (g-z) galaxy colours from F2006; Col. [11]: Integrated (g-z) galaxy colours from  model fit; Col. [12]: Stellar mass to luminosity (B band) ratio from model fit.  }
\begin{tabular}{|c|c|c|c|c|c|c|c|c|c|c|c|}
\hline
\multicolumn{1}{c}{$N_{P2006}$}  &
\multicolumn{1}{c}{VCC}          &
\multicolumn{1} {c} {$N_{blue}$}      &
\multicolumn{1} {c} {$Zs(blue)$} &
\multicolumn{1} {c} {$N_{red}$}      &
\multicolumn{1} {c} {$Zs(red)$}  &
\multicolumn{1} {c} {$\chi^2$}   &
\multicolumn{1} {c} {F}          &
\multicolumn{1} {c} {$g(150)$}   &
\multicolumn{1} {c} {$(g-z)$}    &
\multicolumn{1} {c} {$(g-z)_m$}  &
\multicolumn{1} {c} {$(M/L)_B$}  \\
\multicolumn{1}{c}{(1)}  &
\multicolumn{1}{c}{(2)}          &
\multicolumn{1} {c} {(3)}      &
\multicolumn{1} {c} {(4)} &
\multicolumn{1} {c} {(5)}      &
\multicolumn{1} {c} {(6)}  &
\multicolumn{1} {c} {(7)}   &
\multicolumn{1} {c} {(8)}          &
\multicolumn{1} {c} {(9)}   &
\multicolumn{1} {c} {(10)}    &
\multicolumn{1} {c} {(11)}  &
\multicolumn{1} {c} {(12)}  \\
\hline
1  &  1226  & 188     & 0.037  &    562  &     0.90     & 2.18 &     0.97 &     9.21    &  1.60   &  1.60   &  9.45 \\
2  &  1316  & 388     & 0.034  &   1337  &     0.85     & 3.65 &     0.98 &     9.84    &  1.60   &  1.60   &  9.39 \\
3  &  1978  & 217     & 0.037  &    573  &     1.00     & 0.00 &     0.97 &     9.82    &  1.62   &  1.61   &  9.61 \\
4  &  881   & 122     & 0.024  &    228  &     0.30     & 1.71 &     0.94 &     10.12   &  1.57   &  1.41   &  6.92 \\
5  &  798   & 152     & 0.037  &    354  &     0.45     & 0.40 &     0.95 &     10.00   &  1.38   &  1.48   &  7.81 \\
6  &   763  & 186     & 0.035  &    309  &     0.35     & 2.65 &     0.92 &     9.16    &  1.56   &  1.42   &  7.14 \\
7  &   731  & 204     & 0.030  &    681  &     0.45     & 0.69 &     0.97 &     9.95    &  1.53   &  1.49   &  7.94 \\
8  &   1535 &  70     & 0.018  &    164  &     0.65     & 0.35 &     0.98 &     -----   &  ----   &  1.55   &  8.80 \\
9  &   1903 &  59     & 0.034  &    236  &     0.45     & 0.00 &     0.97 &     10.42   &  1.63   &  1.49   &  7.96 \\
10 &   1632 &  113    & 0.075  &    338  &     0.60     & 2.09 &     0.94 &     10.66   &  1.61   &  1.53   &  8.47 \\
11 &   1231 &  72     & 0.028  &    168  &     0.40     & 1.68 &     0.96 &     10.71   &  1.53   &  1.46   &  7.58 \\
12 &   2095 &  38     & 0.063  &     88  &     0.30     & 0.00 &     0.90 &     11.71   &  1.44   &  1.40   & 6.87  \\
13 &   1154 &  54     & 0.025  &    126  &     0.30     & 0.78 &     0.95 &     10.89   &  1.54   &  1.41   &  6.97 \\
14 &   1062 &  47     & 0.050  &    106  &     0.60     & 0.88 &     0.95 &     11.06   &  1.53   &  1.52   &  8.42 \\
15 &   2092 &  16     & 0.022  &     64  &     0.35     & 0.42 &     0.98 &     11.13   &  1.50   &  1.45   &  7.43 \\
16 &   369  &  51     & 0.031  &    119  &     0.45     & 0.00 &     0.96 &     12.22   &  1.57   &  1.48   &  7.84 \\
17 &   759  &  68     & 0.031  &     92  &     0.40     & 0.72 &     0.92 &     11.34   &  1.54   &  1.44   &  7.37 \\
18 &   1692 &  60     & 0.025  &     60  &     0.50     & 1.03 &     0.92 &     11.65   &  1.53   &  1.47   &  7.79 \\
19 &   1030 &  50     & 0.022  &    116  &     0.45     & 0.38 &     0.97 &     -----   &  ----   &  1.48   &  7.9  \\
20 &   2000 &  102    & 0.025  &     83  &     0.25     & 1.09 &     0.86 &     11.71   &  1.51   &  1.34   &  6.24 \\
21 &   685  &  78     & 0.031  &     78  &     0.50     & 1.24 &     0.91 &     11.59   &  1.55   &  1.46   &  7.71 \\
22 &   1664 &  26     & 0.025  &    104  &     0.35     & 1.01 &     0.97 &     11.63   &  1.54   &  1.45   &  7.42 \\
23 &   654  &  24     & 0.031  &     16  &     0.50     & 0.90 &     0.87 &     11.98   &  1.45   &  1.44   &  7.44 \\
24 &   944  &  40     & 0.031  &     40  &     0.45     & 1.26 &     0.90 &     11.81   &  1.48   &  1.45   &  7.47 \\
25 &   1938 &  49     & 0.021  &     40  &     0.20     & 0.27 &     0.85 &     11.77   &  1.49   &  1.31   &  5.90 \\
26 &   1279 &  55     & 0.020  &     75  &     0.25     & 0.86 &     0.92 &     11.96   &  1.46   &  1.37   &  6.50 \\
28 &   355  &  24     & 0.018  &     26  &     0.50     & 0.27 &     0.94 &     12.08   &  1.52   &  1.48   &  7.93 \\
29 &   1619 &  12     & 0.018  &     43  &     0.30     & 0.13 &     0.97 &     12.18   &  1.42   &  1.42   &  7.09 \\
30 &   1883 &  30     & 0.031  &     45  &     0.25     & 0.63 &     0.90 &     11.76   &  1.32   &  1.36   &  6.45 \\
31 &   1242 &  38     & 0.031  &     72  &     0.30     & 0.97 &     0.93 &     12.19   &  1.46   &  1.40   &  6.87 \\
34 &   778  &  22     & 0.025  &     38  &     0.30     & 0.42 &     0.93 &     12.67   &  1.47   &  1.40   &  6.88 \\
35 &   1321 &  22     & 0.037  &     22  &     0.35     & 0.09 &     0.87 &     12.28   &  1.33   &  1.40   &  6.88 \\
36 &   828  &  35     & 0.017  &     35  &     0.25     & 0.19 &     0.91 &     12.71   &  1.48   &  1.36   &  6.42 \\
37 &   1250 &  22     & 0.025  &     22  &     0.08     & 0.55 &     0.75 &     12.40   &  1.21   &  1.19   &  4.79 \\
38 &   1630 &  10     & 0.030  &     24  &     0.50     & 1.14 &     0.96 &     12.54   &  1.50   &  1.50   &  8.09 \\
40 &   1025 &  65     & 0.044  &     35  &     0.30     & 0.00 &     0.74 &     12.61   &  1.43   &  1.33   &  6.17 \\
41 &   1303 &  41     & 0.024  &     14  &     0.50     & 0.63 &     0.80 &     12.68   &  1.42   &  1.39   &  6.94 \\
42 &   1913 &  22     & 0.031  &     33  &     0.60     & 0.38 &     0.94 &     12.90   &  1.43   &  1.51   &  8.35 \\
44 &   1125 &  32     & 0.016  &     18  &     0.20     & 0.55 &     0.83 &     12.85   &  1.42   &  1.29   &  5.78 \\
45 &   1475 &  38     & 0.031  &     38  &     0.15     & 1.31 &     0.80 &     13.06   &  1.35   &  1.27   &  5.49 \\
46 &   1178 &  32     & 0.015  &     48  &     0.30     & 0.86 &     0.95 &     13.06   &  1.47   &  1.41   &  6.95 \\
47 &   1283 &  29     & 0.022  &     26  &     0.30     & 0.47 &     0.89 &     13.02   &  1.47   &  1.38   &  6.66 \\
48 &   1261 &  40     & 0.088  &      0  &     -----    & 0.46 &     0.00 &     13.29   &  1.20   &  1.22   &  4.96 \\
49 &   698  &  25     & 0.015  &     85  &     0.10     & ---- &     0.95 &     13.04   &  1.38   &  1.25   &  5.27 \\
50 &   1422 &  25     & 0.101  &      0  &     -----    & 0.29 &     0.00 &     13.60   &  1.27   &  1.24   &  5.11 \\
53 &   9    &  25     & 0.044  &      0  &     -----    & 0.73 &     0.00 &     13.81   &  1.15   &  1.14   &  4.36 \\
57 &   856  &  35     & 0.056  &     0   &     -----    & 0.49 &     0.00 &     14.25   &  1.22   &  1.17   &  4.54 \\
60 &   1087 &  45     & 0.044  &      0  &     -----    & 0.63 &     0.00 &     13.87   &  1.29   &  1.14   &  4.36 \\
62 &   1861 &  35     & 0.094  &      0  &     -----    & 1.05 &     0.00 &     14.22   &  1.33   &  1.23   &  5.04 \\
63 &   1431 &  65     & 0.075  &      0  &     0.05     & 0.60 &     0.00 &     14.12   &  1.24   &  1.20   &  4.80 \\
63 &   543  &  20     & 0.031  &      0  &     -----    & 0.50 &     0.00 &     14.12   &  1.24   &  1.10   &  4.15 \\
67 &   1833 &  20     & 0.101  &      0  &     -----    & 0.53 &     0.00 &     14.46   &  1.21   &  1.24   &  5.11 \\
68 &   437  &  25     & 0.018  &      0  &     -----    & 0.91 &     0.00 &     13.82   &  1.26   &  1.05   &  3.93 \\
71 &   200  &  15     & 0.018  &      0  &     -----    & 0.87 &     0.00 &     14.79   &  1.21   &  1.05   &  3.93 \\
73 &   21   &  15     & 0.018  &      0  &     -----    & 0.74 &     0.00 &     14.79   &  0.82   &  1.05   &  3.93 \\
78 &   1545 &  30     & 0.017  &     20  &     0.15     & 0.76 &     0.82 &     14.77   &  1.34   &  1.26   &  5.43 \\
81 &   1075 &  20     & 0.025  &      0  &     -----    & 0.65 &     0.00 &     14.96   &  1.20   &  1.08   &  4.04 \\
84 &   1440 &  21     & 0.018  &      9  &     0.50     & 0.00 &     0.87 &     14.67   &  1.26   &  1.42   &  7.33 \\
\hline
\end{tabular}
\label{sample}
\end{table*}

\begin{table*}
\caption{Table 1 -- Continued. }
\begin{tabular}{|c|c|c|c|c|c|c|c|c|c|c|c|}
\hline
\multicolumn{1}{c}{$N_{P2006}$}  &
\multicolumn{1}{c}{VCC}          &
\multicolumn{1} {c} {$N_{blue}$}      &
\multicolumn{1} {c} {$Zs(blue)$} &
\multicolumn{1} {c} {$N_{red}$}      &
\multicolumn{1} {c} {$Zs(red)$}  &
\multicolumn{1} {c} {$\chi^2$}   &
\multicolumn{1} {c} {F}          &
\multicolumn{1} {c} {$g(150)$}   &
\multicolumn{1} {c} {$(g-z)$}    &
\multicolumn{1} {c} {$(g-z)_m$}  &
\multicolumn{1} {c} {$(M/L)_B$}  \\
\multicolumn{1}{c}{(1)}  &
\multicolumn{1}{c}{(2)}          &
\multicolumn{1} {c} {(3)}      &
\multicolumn{1} {c} {(4)} &
\multicolumn{1} {c} {(5)}      &
\multicolumn{1} {c} {(6)}  &
\multicolumn{1} {c} {(7)}   &
\multicolumn{1} {c} {(8)}          &
\multicolumn{1} {c} {(9)}   &
\multicolumn{1} {c} {(10)}    &
\multicolumn{1} {c} {(11)}  &
\multicolumn{1} {c} {(12)}  \\
\hline

85  &   230  &  20     & 0.018  &      0  &     -----    & 0.92 &     0.00 &     15.45   &  1.17   &  1.05   &  3.93 \\
89 &   1828 &  20     & 0.031  &      0  &     -----    & 0.50 &     0.00 &     15.09   &  1.25   &  1.10   &  4.15 \\
91 &   1407 &  45     & 0.101  &      0  &     -----    & 0.60 &     0.00 &     15.03   &  1.23   &  1.24   &  5.11 \\
95 &   1539 &  35     & 0.063  &      0  &     -----    & 0.00 &     0.00 &     15.15   &  1.21   &  1.18   &  4.63 \\
96 &   1185 &  25     & 0.018  &      0  &     -----    & 0.66 &     0.00 &     15.24   &  1.28   &  1.05   &  3.93 \\
\hline
\end{tabular}
\end{table*}

\begin{figure*}
\resizebox{0.3\hsize}{!}{\includegraphics{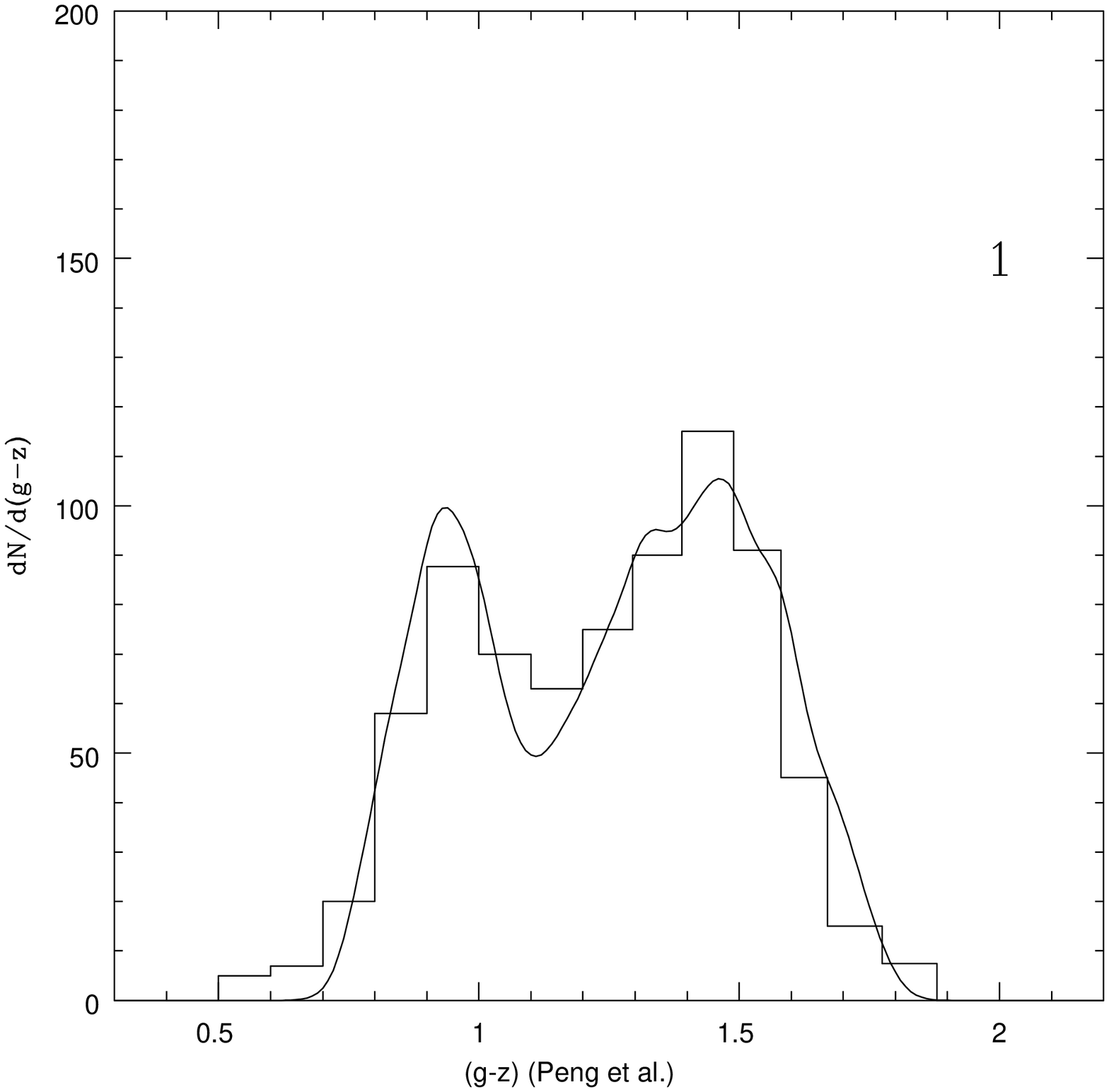}}
\resizebox{0.3\hsize}{!}{\includegraphics{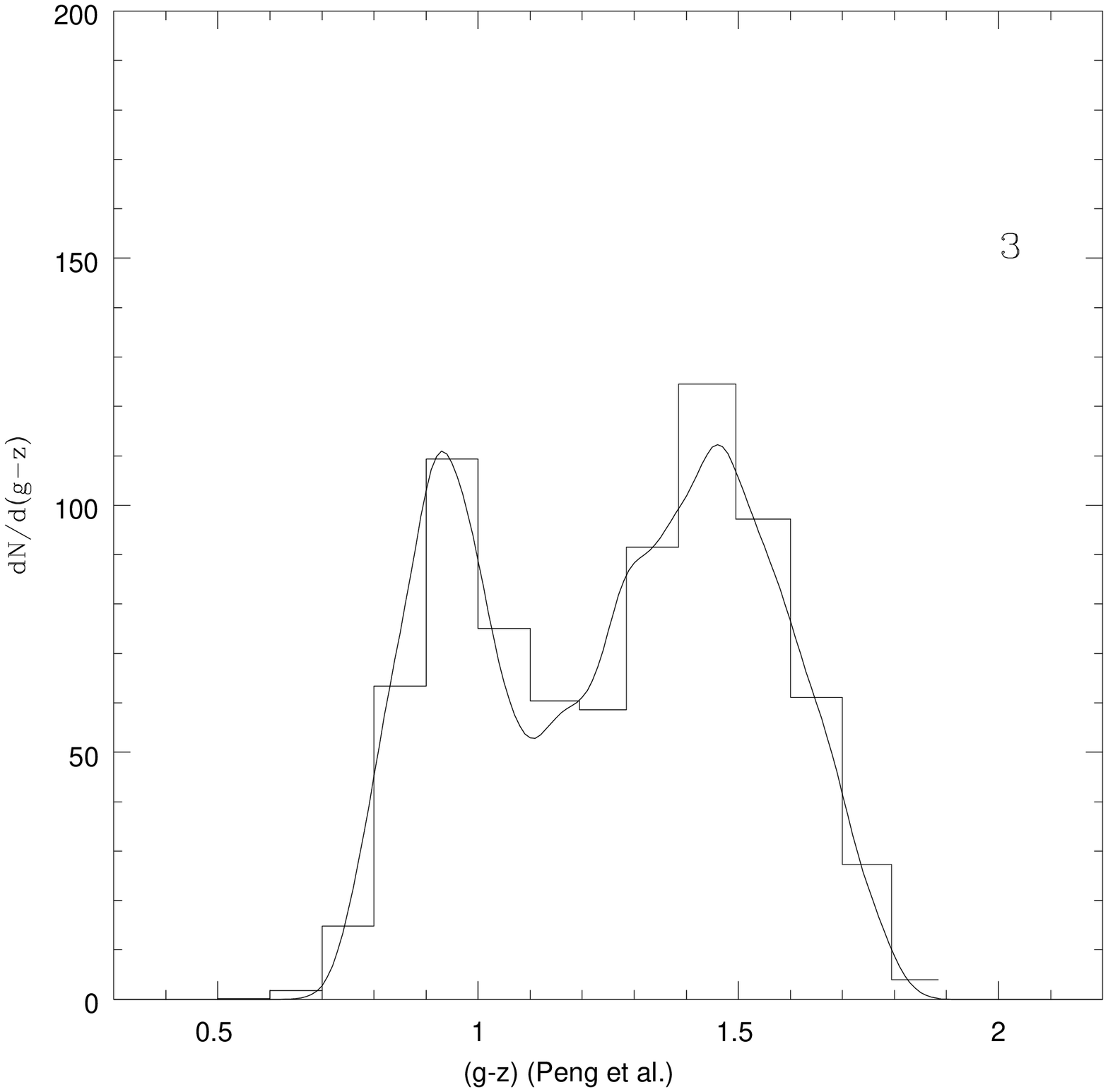}}
\resizebox{0.3\hsize}{!}{\includegraphics{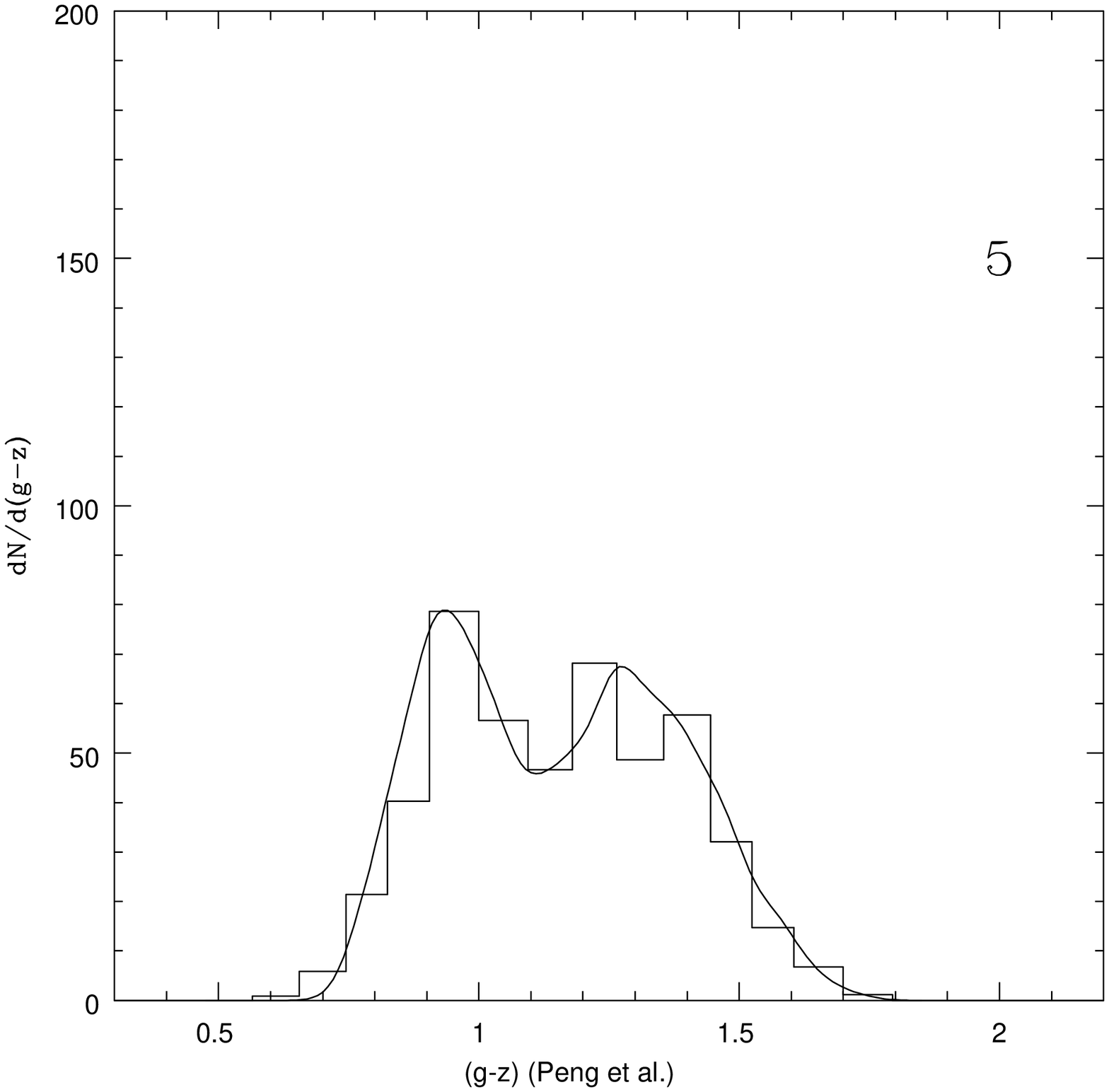}}
\resizebox{0.3\hsize}{!}{\includegraphics{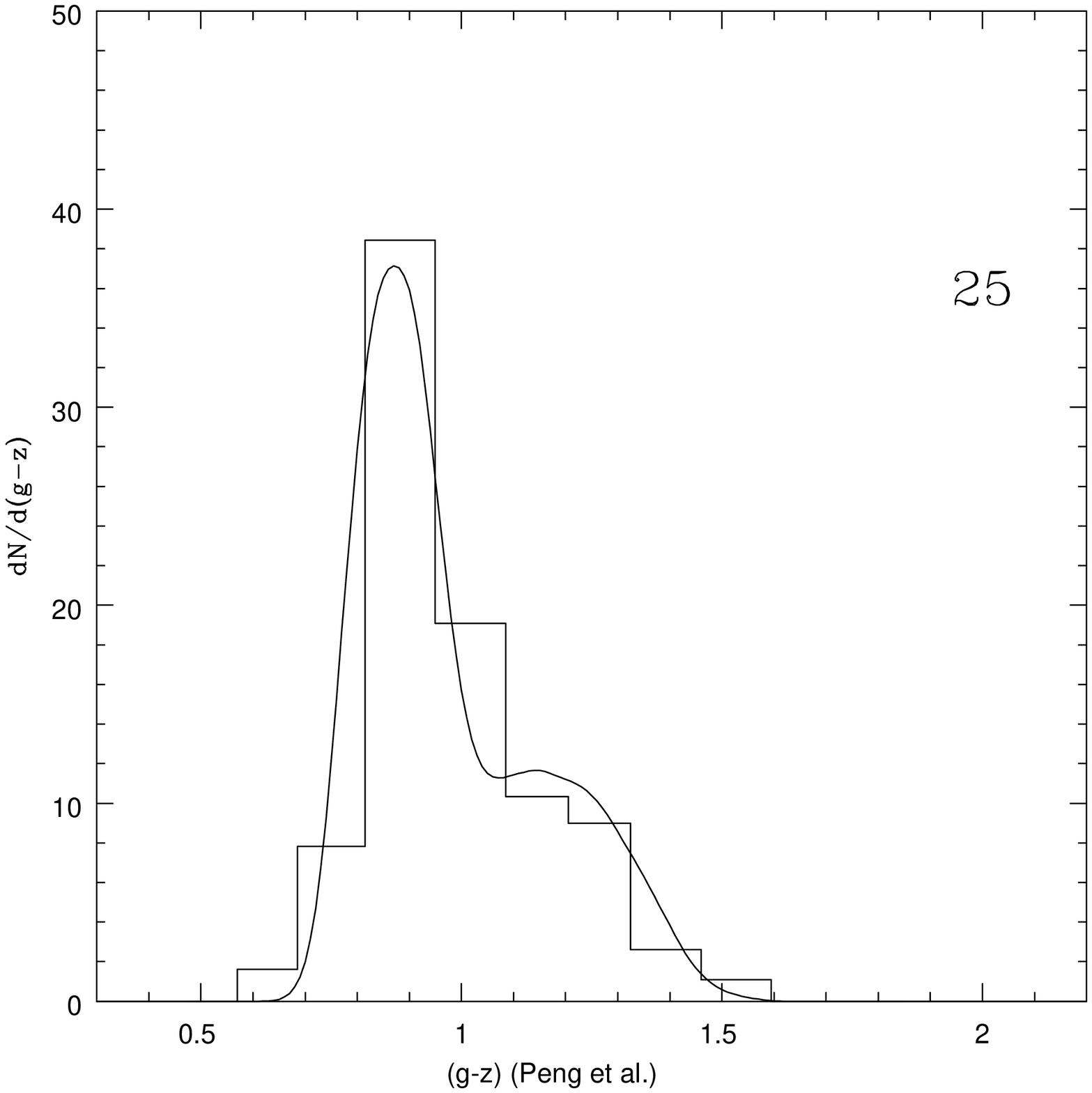}}
\resizebox{0.3\hsize}{!}{\includegraphics{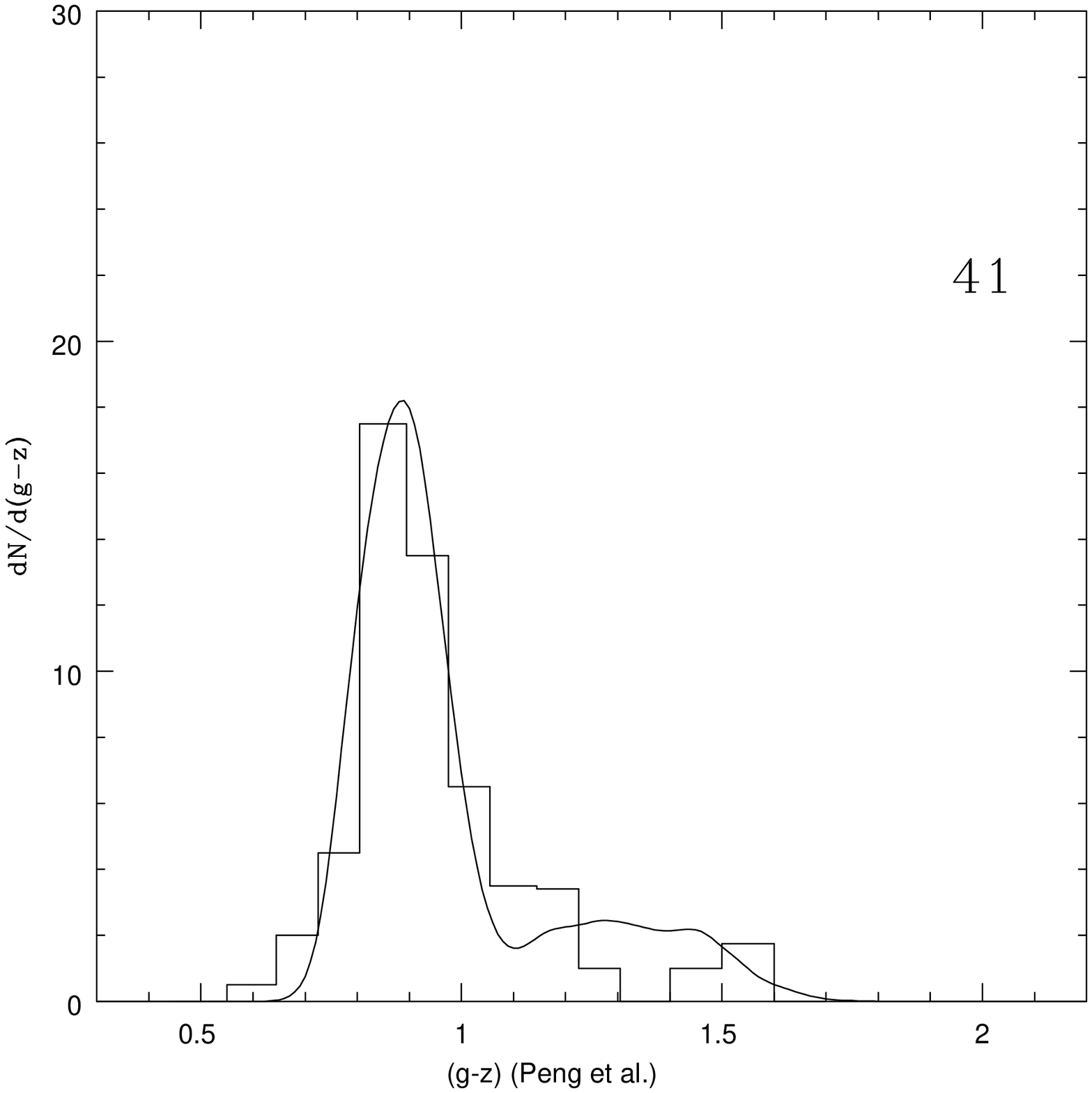}}
\resizebox{0.3\hsize}{!}{\includegraphics{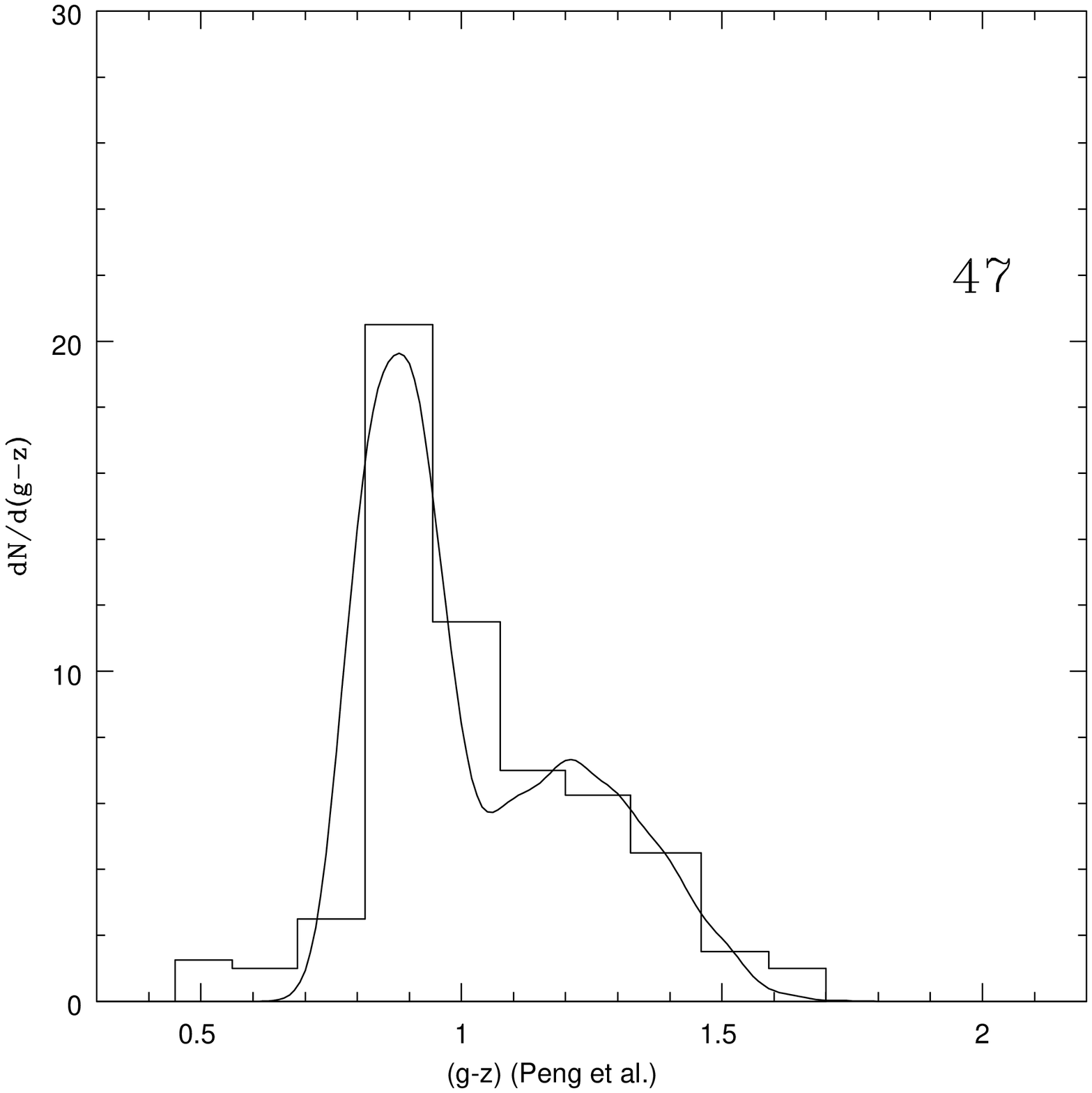}}
\resizebox{0.3\hsize}{!}{\includegraphics{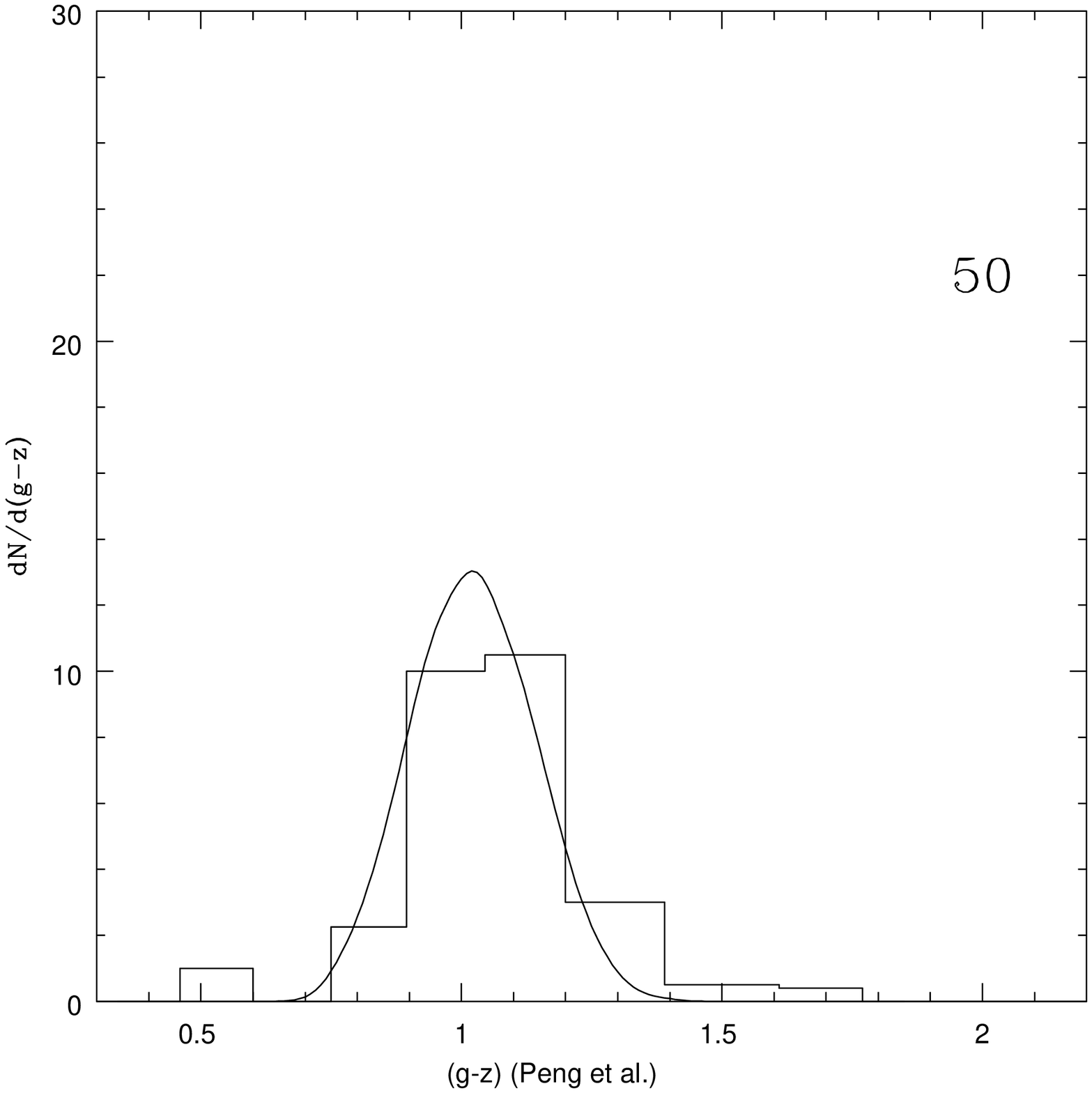}}
\resizebox{0.3\hsize}{!}{\includegraphics{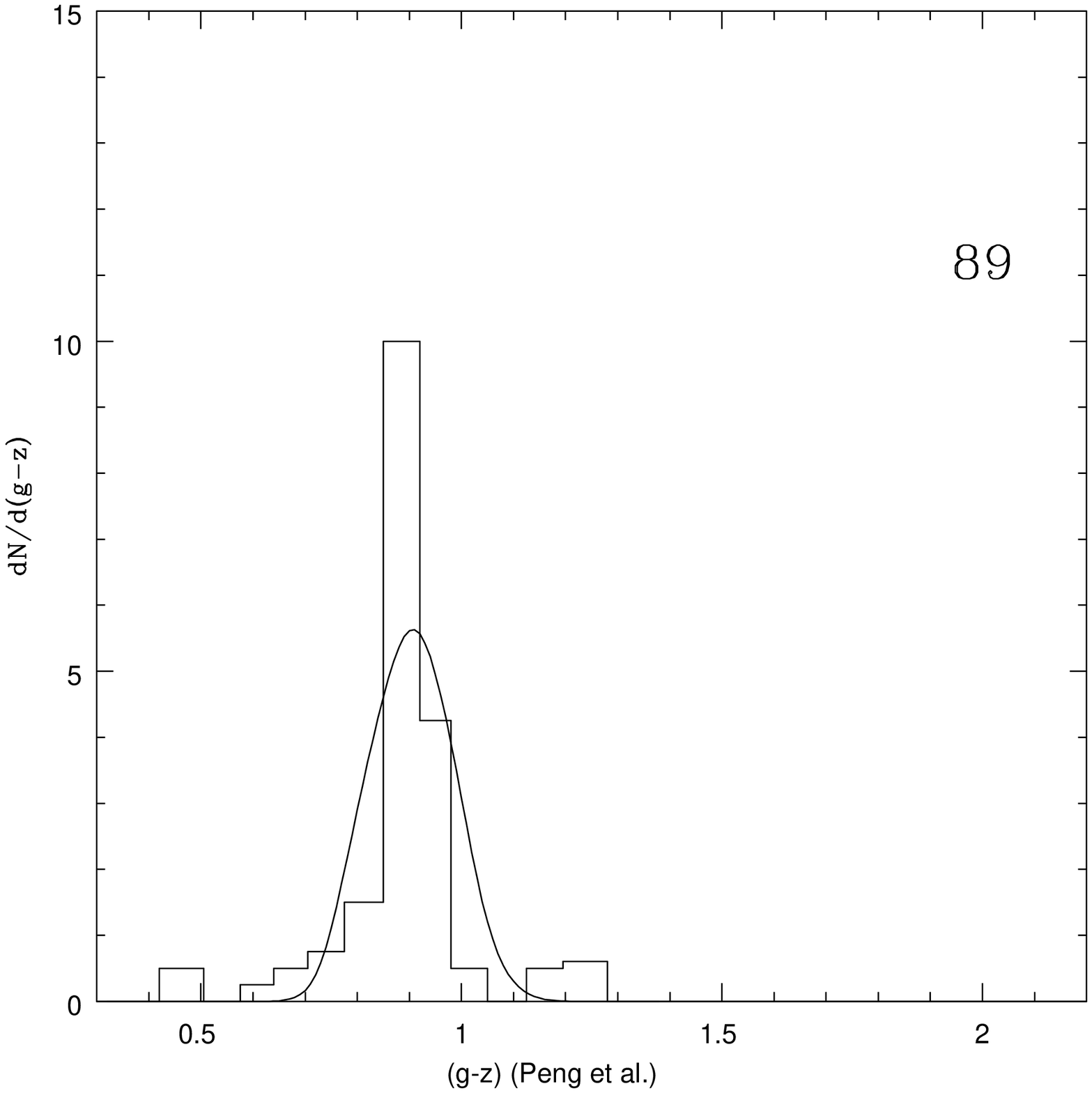}}
\resizebox{0.3\hsize}{!}{\includegraphics{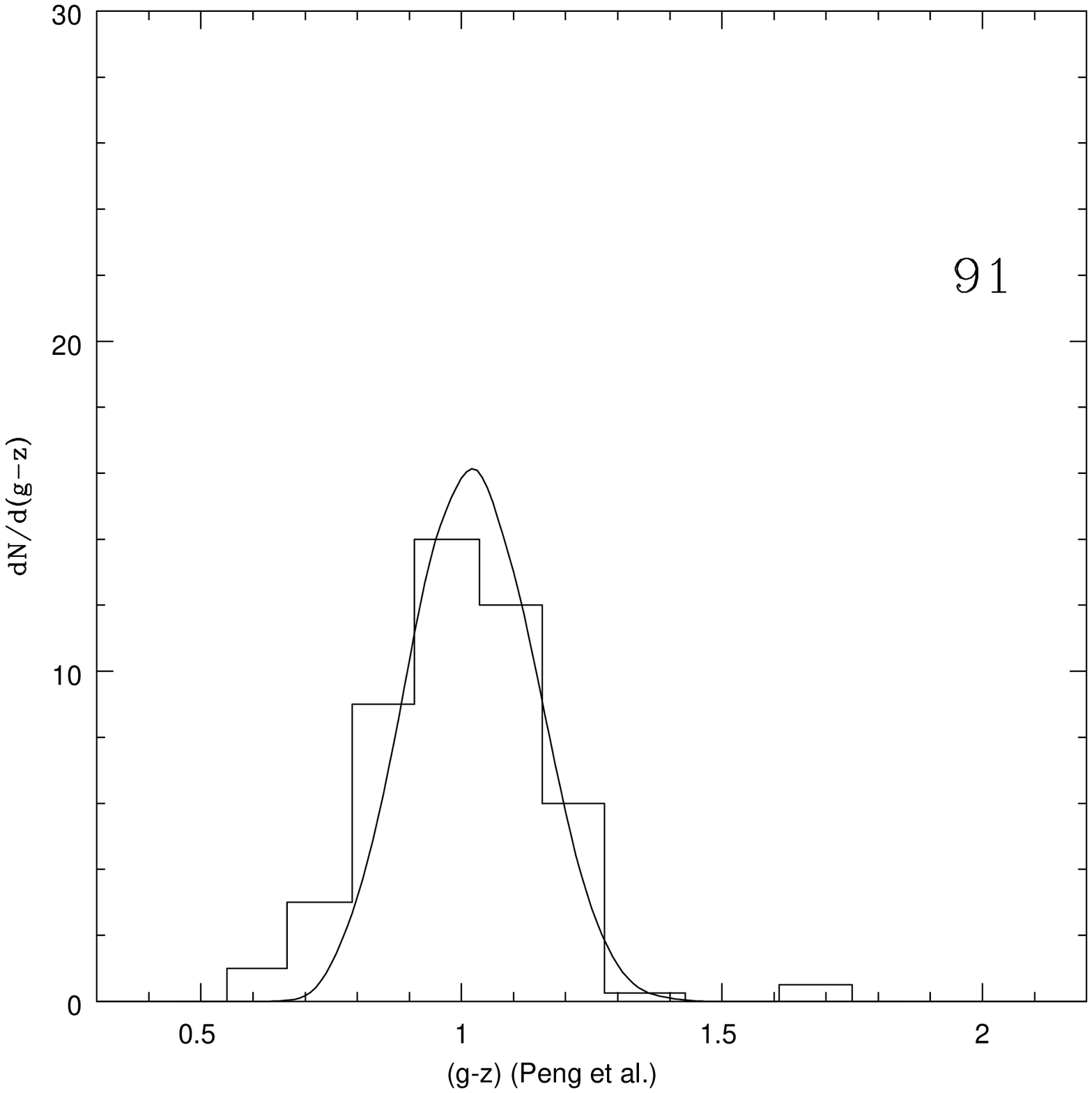}}
\caption{ Sample GC colour histogram fits. Upper row: VCC 1226, VCC1978, VCC 798; middle row 
VCC 1938, VCC 1303, VCC 1283;lower row: VCC 1422, VCC 1828, VCC 1407. Upper right numbers are
the galaxy identifications in Peng et al. (2006).}  
\label{sample_fits}
\end{figure*}

  A sample of histogram fits are shown in Figure \ref{sample_fits} for some 
  representative galaxies with different numbers of GC (upper row: galaxies 
  with more than 250 GC; middle row: with 50 to 100 GC; lower row: with 
  less than 50 GC). In this figure the fits are
  smoothed histograms adopting a Gaussian colour kernel of 0.03 mags on 
  the model GC colours. In turn, the histogram fits  for the 63 galaxies 
  listed in Table\,\ref{sample}, including the associated statistical counting 
  uncertainties for each colour bin are displayed in the Appendix.

  Figure \ref{sample_fits} shows the same trend noted already by P2006. in the
  sense that GC colour bi-modality is clearly present in the brightest galaxies and
  becomes less evident, or is directly absent, in the faintest ones.

\section{The $\gamma$ and  $\delta$  parameters}

\label{GAMMADELTA}

  This paragraph revises the values of the  $\gamma$ and $\delta$ parameters
  given in FFG07 on the basis of the following arguments:

  a) Those authors used $[Z/H]$ instead of $[Fe/H]_{zw}$ (as stated in the paper)
  as the argument of the adopted colour-metallicity relation used to obtain
  integrated model colours. That mistake is corrected in this work.
  As in FFG07, $[Z/H]$ is on the \citet{TMK04} chemical abundance scale and 
  connected  with $[Fe/H]_{zw}$  through $[Z/H] = [Fe/H]_{zw} + 0.131$ as derived by
  \citet{MPF07}.

  b) The $\bf \delta$ parameter was determined in FFG07 only trough the fit to the observed
  brightness profiles of NGC 1399 and NGC 4486. However, that parameter also
  controls the resulting galactocentric colour gradient. In this paper, we
  attempted a simultaneous fit to both the brightness and colour gradients,
  as given in \citet{Mi00} for these galaxies, aiming at minimizing the
  orthogonal sum of the rms of the brightness and colour fits.

  c) $\bf \gamma$ and $\bf\delta$ were also derived for NGC 1427 using data from
  \citet {FO01}.
  This galaxy was observed adopting the $(C-T_1)$ index and using the
  same reduction and calibration procedures reported by FFG07.

  The revised parameters for the three galaxies are listed in Table\,\ref{gamma_delta}.
  In particular, the $\bf\delta$ parameters seem comparable (within the
  uncertainties) then suggesting that a similar pattern
  connects GC and stellar haloes in the three systems. We note that the
  average value, $\bf \delta$ = 1.75  that we adopt hereafter, is higher than those
  given in FFG07 ($\bf\delta$ = 1.10 and 1.20 ) for NGC 1399 and NGC 4486,
  respectively, after correcting the mistake noted in the first item of this paragraph.

\begin{table}
\caption{$\gamma$ and $\delta$ parameters for NGC 1399, NGC 1427 and NGC 4486.}
\label{gamma_delta}
\begin{tabular}{|l|c|c|c|}
\hline
\multicolumn{1}{c}{} &
\multicolumn{1}{c}{NGC 1399} &
\multicolumn{1} {c} {NGC1427}  &
\multicolumn{1} {l} {NGC 4486} \\
\hline
 $\gamma~$ ($10^{-8}$ units)    &    0.73  $\pm$ 0.05  &   0.65 $\pm$ 0.03  & 0.75 $\pm$ 0.03 \\
 $\delta$    &    1.60  $\pm$ 0.15  &   1.90 $\pm$ 0.15  & 1.75 $\pm$ 0.10 \\
\hline
\end{tabular}
\end{table}

\section{Integrated colours, stellar mass-to-light ratios and galaxy masses}
\label{MODEL}

   The model assumptions, i.e.:
\begin{equation}
dN/dZ = (N/Zs) \exp(-Z/Zs)\
\end{equation}

\noindent where N is the number of globular clusters in the blue or red families
   and $Zs$ their respective abundance scales, and:

\begin{equation}
dN/dM_*(Z) = {\gamma} \exp({-{\delta~[Z/H])}}\               
\end{equation}

\noindent imply that the amount of diffuse stellar mass with chemical abundance
   $Z$ will be given by:

\begin{eqnarray}
M_*(Z)=\gamma^{-1} \exp(\delta~[Z/H]) \big[\frac{N_{blue}}{Zs(blue)}\exp(-Z/Zs(blue)) +  \nonumber  \\
\frac{N_{red}}{Zs(red)}\exp(-Z/Zs(red))\big]
\end{eqnarray}

   Both $N_{blue}$ and $N_{red}$ have distinct behaviours as a function of galactocentric
   radius which can be used to derive surface brightness profiles as shown
   in FFG07. However,
   in this work, they represent the total number of blue or red GC within the
   ACS fields.

   The B luminosity associated to $M_*(Z)$ was estimated by adopting a
   reference mass-to-luminosity ratio:

\begin{equation}
(M/L)_{B} = 3.71 + ([Z/H] + 2.0)^{2.5} 
\end{equation}

\noindent with $(M/L)_{B} = 3.71 $ for $[Z/H] < -2$ and $(M/L)_{B} = 13.6 $ for
   $[Z/H] > 0.5$, which gives a good approximation (within $\pm 7$ percent) to 
   the SSP models by \citet{Wo94} for an age of 12 Gy and a Salpeter luminosity
   function. A comparison with other models gives an idea about the
   uncertainty of this ratio. For example, models by \citet{MA04}, for
   the same age and luminosity function, shows an overall
   agreement within 10 percent with Worthey's except at the lowest
   abundance where the ratio is about 24 percent larger.

   The total stellar  mass, integrated $B$ luminosity and composite
   $(M/L)_{B}$ ratio can be found by integrating $M_*(Z)~dZ$ and $M_*(Z)~(M/L)^{-1}_{B}~dZ$,
   respectively, where we adopted $Z_{l}=0.0035~Z_{\odot}$ and $Z_{u}=4.0~Z_{\odot}$ as 
   lower and upper integration limits.

    In what follows, we tentatively set $\bf \delta$  = 1.75  as representative for all galaxies
    (see paragraph \ref{GAMMADELTA}). With this assumption, the integrated mass-to-light ratios, as
    well as the total mass fraction in a given stellar population,
    will only depend on the $N_{blue}$, $Zs(blue)$, $N_{red}$ and $Zs(red)$ values 
    obtained from the GC
    colour histogram fits and are independent of the $\bf \gamma$ parameter. 
    This also holds for the integrated galaxy colours that, for
    each $M(Z)$, depend only on the chemical abundance through the adopted
    colour-metallicity relation and on the value of $\bf \delta$, which
    determines the amount of stellar mass with a given abundance. 

    Table\,\ref{sample} lists 
    the stellar mass fraction associated
    to the red GC population, {\bf F}, the integrated $g(150)$ magnitude,
    the $(g-z)$ galaxy colours from F2006, the $(g-z)$ colours derived from the
    model and, finally, the inferred stellar mass to B luminosity ratio.

    In the case of M87, the adoption of the parameters given in Table\,\ref{sample}
    leads to  $(g-z)$ = 1.55 , which is 0.05 bluer than the value quoted by
    F2006. This difference can be removed by increasing the $Zs(red)$ scale by 1.25,
    i.e., a fraction of the stellar mass associated with the red
    GC, in the inner 150 arcsecs, seem to have formed with $Z$ values
    somewhat higher than in these clusters. We note that this is not the
    case at galactocentric radii larger than 120 arcsecs where the
    integrated colours of both NGC 1399 and NGC 4486 can be matched with
    the $Zs(red)$ derived from the histogram fits (as shown in FFG07).

    An argument in the same sense can be found in \citet{BEA08} who
    show that, in the {\bf inner} regions of the resolved galaxy NGC 5128,
    field stars have upper $[Z/H]$ values some 0.2 dex higher than those
    of the red GC. On the basis of these considerations we
    adopted a chemical abundance spread equal 1.25 $Zs(red)$ in the estimate of
    the integrated colours of all our sample galaxies.

\begin{figure*}
\resizebox{0.4\hsize}{!}{\includegraphics{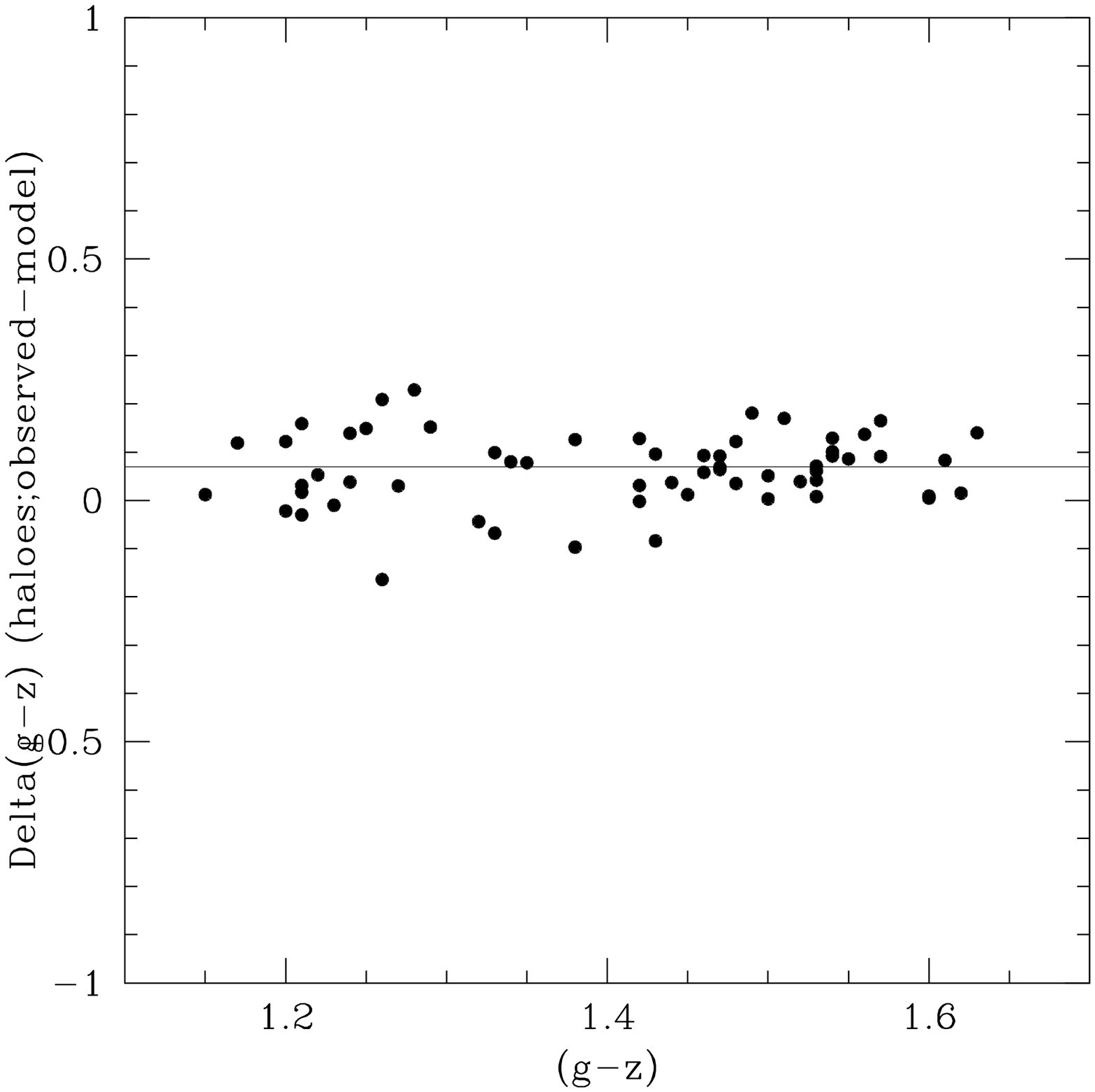}}
\resizebox{0.4\hsize}{!}{\includegraphics{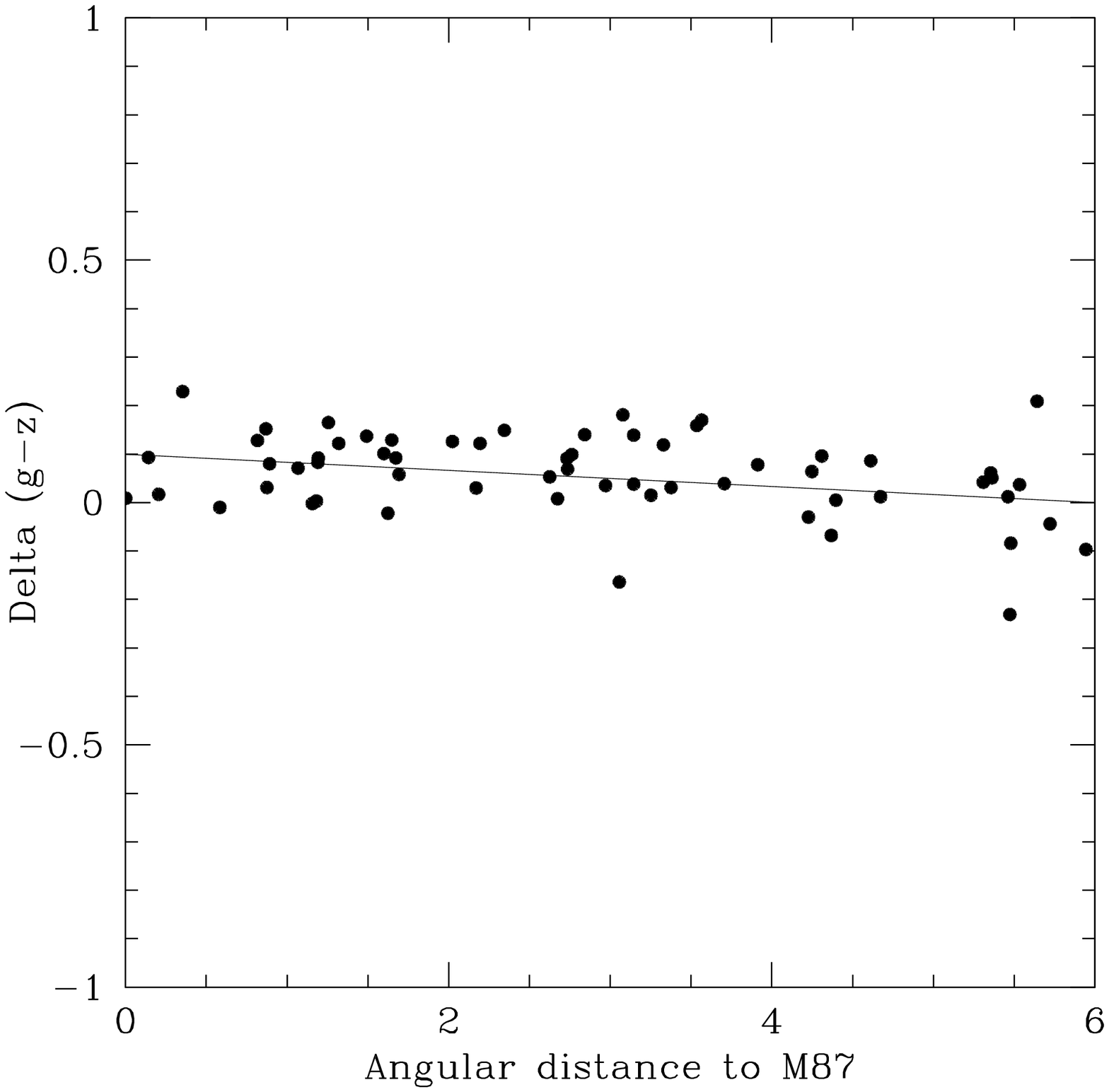}}
\caption{Left panel:$(g-z)$ colour differences between the observed halo colours (from 
   Ferrarese et al. 2006) and those obtained from modelling as described
   in this paper. 
   Right panel: Stellar haloes $(g-z)$ colour residual (observed minus model) as a 
   function of the angular distance to M87. The straight line suggests a marginal
   trend of colour residuals with position within the Virgo cluster.}
\label{residuals1}
\end{figure*}

    The residual colours (observed minus model) as a function of
    the integrated $(g-z)$ colours given by F2006 are shown in Figure 
    \ref{residuals1} (left panel). The expected impact on the calculated $(g-z)$
    colours, arising the uncertainties of the chemical abundance scales
    referred to in paragraph \ref{HISTOFITS}, ranges from 0.02 to 0.06 mags.
    The larger dispersion in this diagram then probably arises in observational 
    errors that become more important for the bluer (and fainter) galaxies.


     This diagram shows that the model is able to approximate the
     observed galaxy colours although
     an average $(g-z)$ residual of 0.07 mags remains. This residual,
     as shown in Figure \ref{residuals1} (right panel), correlates marginally with the
     distance to  the centre of the Virgo cluster (that we tentatively identify with
     NGC 4486) in the sense that, inner galaxies, appear slightly redder than the colours
     predicted by the model. We note that galaxies number 73 (VCC 21; not included
     in Figure 5a) and number 84 (VCC 1440), exhibit $(g-z)$ colours that are
     significantly bluer than those inferred from their GC and possibly denoting the
     presence of younger stellar populations. 

\begin{figure*}
\resizebox{0.4\hsize}{!}{\includegraphics{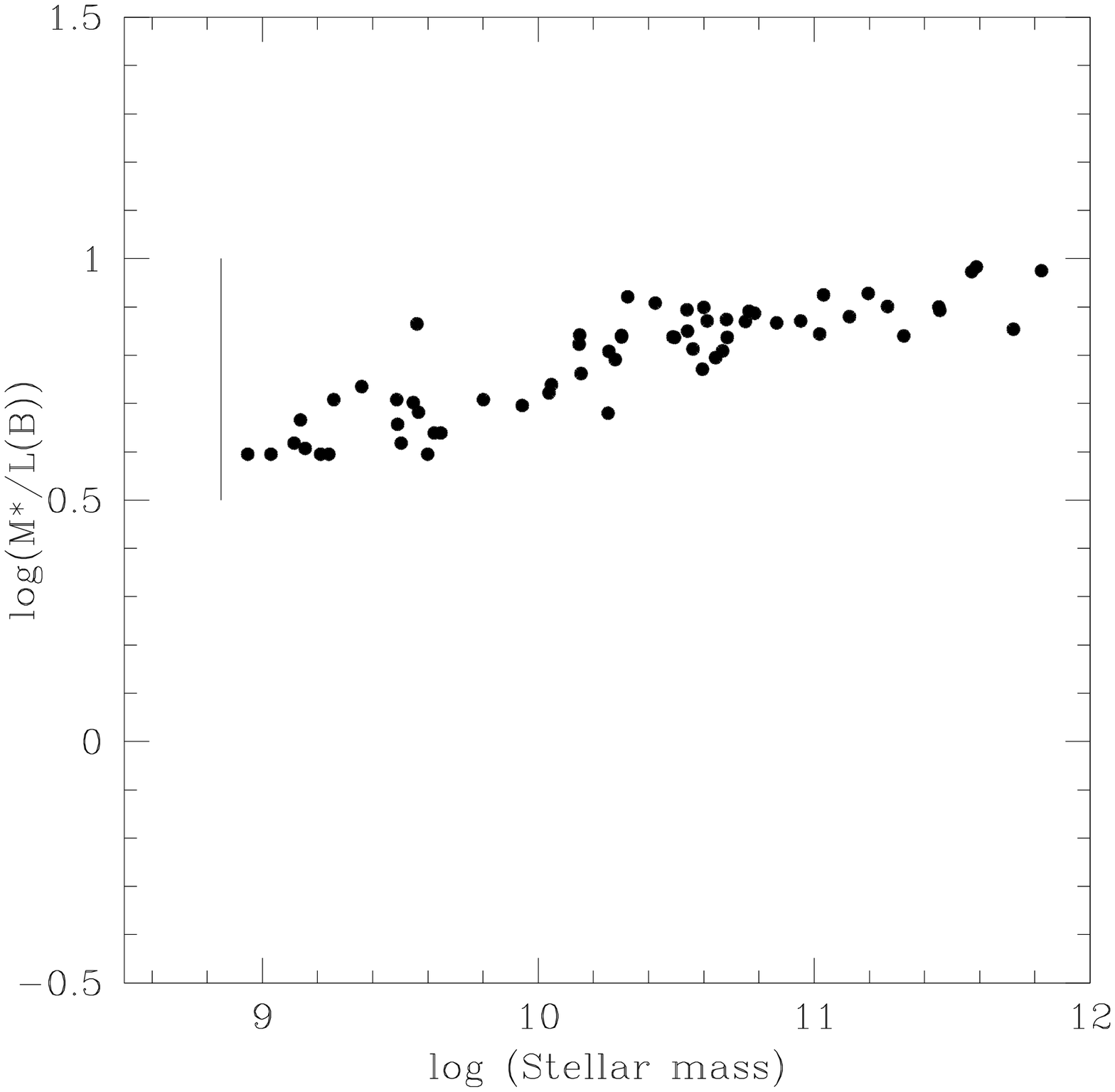}}
\resizebox{0.4\hsize}{!}{\includegraphics{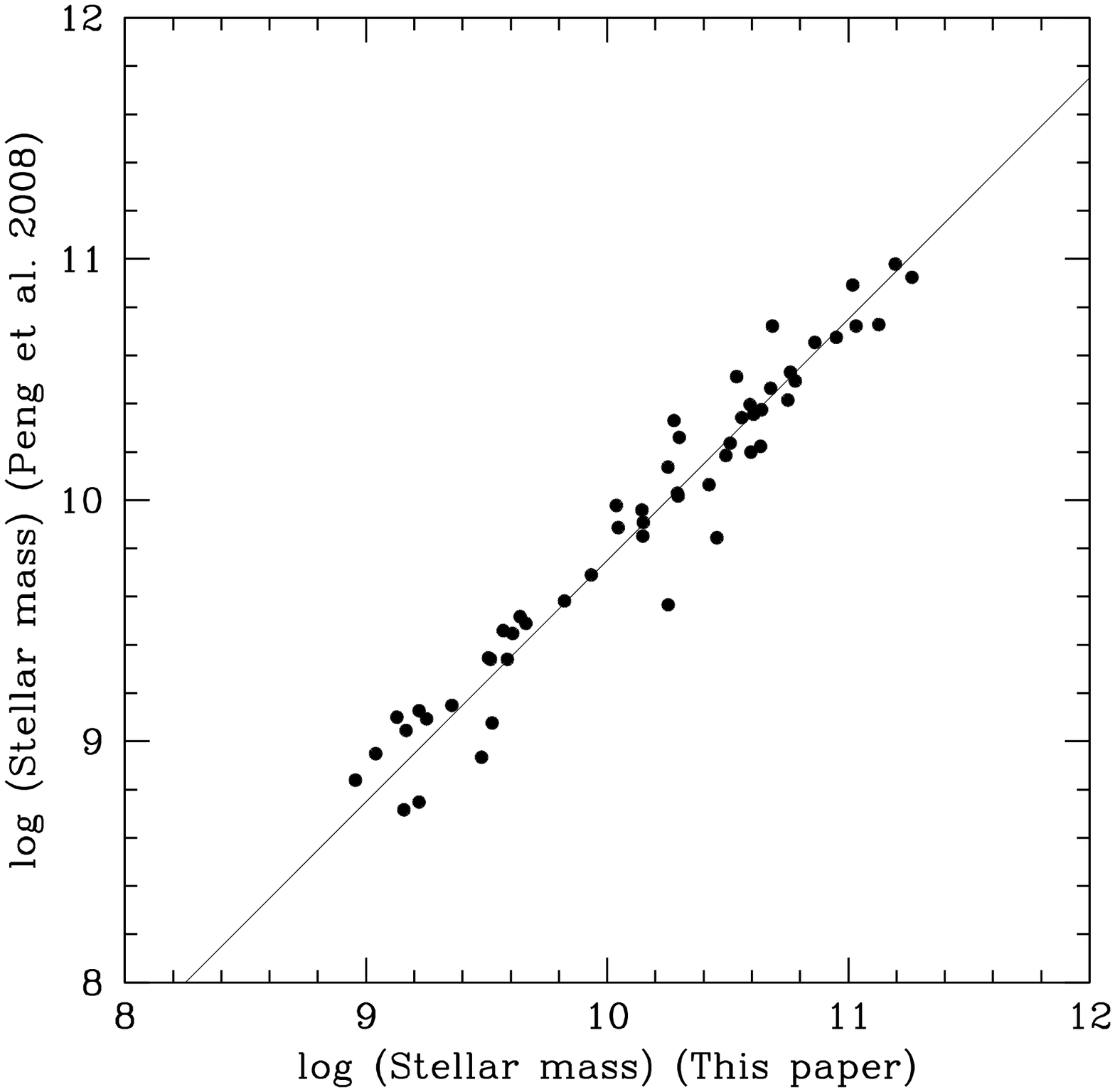}}
\caption{Left panel:Stellar mass to Blue luminosity ratio as a function of total stellar
 mass derived in this paper for 63 galaxies in the ACS Virgo Survey. Right panel:
Comparison between the stellar masses derived in this paper with those
   given by Peng et al. (2006). The straight line (with unity slope) is the approximation
   given in the text.}
\label{masslum}
\end{figure*}

     Stellar mass-to-B luminosity ratios as a function of stellar mass
     are depicted in Figure \ref{masslum} (left panel). The vertical line in this diagram 
     shows the range in $(M/L)_{B}$  obtained by \citet{NAP05}. These authors present a 
     comparison between ratios derived through dynamical and SSP models that, according to
     the last diagram, are in very good agreement with the values obtained in this paper.
     We stress that the relation between the $(M/L)_{B}$ values and the blue absolute
     magnitude of the galaxies (listed in Table 1 and 3, respectively) show good agreement
     with the analysis of the SDSS data given by \cite{KAUFF03}. In particular, those
     values define an upper envelope to their relation suggesting that the VCC galaxies
     in our sample are in fact the oldest for a given galaxy brightness.


     Total masses within {\bf a}=150 arcsecs were derived
     by transforming $g(150)$ magnitudes to $B$ absolute magnitudes (as explained in paragraph
     \ref{PHOTSC}), adopting a distance modulus to Virgo, $(V-M_V)_0 = 31.1$
     from \citet{Ton01}, and the  $(M/L)_{B}$ ratios given in Table\,\ref{parametros} (column 6). This table
     also lists the absolute magnitudes $M_B$ and angular distance of each galaxy
     to NGC 4486 (in degrees). 

     The derived stellar masses $M_*$ are compared with those given by \citealt{Peng08}
     in Figure \ref{masslum} (right panel). This comparison excludes the eight 
     brightest galaxies since, being extended objects, they have a considerable
     fraction of their stellar masses beyond {\bf a}=150 arcsecs. The straight line
\begin{equation}
log(M_*)_{\rm Peng~et~al.}=1.0~log(M_*)-0.25
\end{equation}
     indicates that our masses are a factor about 1.8 larger. This difference arises since
     our reference mass to luminosity ratio (eqn. 11) adopts a Salpeter stellar mass
     distribution while \citealt{Peng08} models stand on that derived by \citealt{CHAB2003}.
     The GC 
     formation efficiency discussed later, however, will keep the same dependence on stellar 
     mass as the slope of the relation between both mass estimates is not significantly
     different from unity.

\begin{table*}
\caption{Chemical abundances, structural paramteters and globular cluster
 formation efficiency parameters for Virgo ACS galaxies.}
\begin{tabular}{|c|c|c|c|c|c|c|c|c|}
\hline
\multicolumn{1}{c}{$N_{P2006}$}     &
\multicolumn{1}{c}{VCC}             &
\multicolumn{1} {c} {$M_B(150)$}    &
\multicolumn{1} {c} {$r(degrees)$}        &
\multicolumn{1} {c} {$[Z/H]$}       &
\multicolumn{1} {c} {$log(M_*)$}     &
\multicolumn{1} {c} {$log(\sigma)$}  &
\multicolumn{1} {c} {$log(r_e)$}     &
\multicolumn{1} {c} {$log(t)$}  \\
\hline
    1 &   1226 &   -21.64 &   4.4  &   0.05  &  11.82  &   8.58  &   1.22  &   0.05 \\
    2 &   1316 &   -21.02 &   0.0  &   0.04  &  11.57  &   8.53  &   1.12  &   0.67 \\
    3 &   1978 &   -21.03 &   3.3  &   0.07  &  11.59  &   9.00  &   0.90  &   0.31 \\
    4 &    881 &   -20.73 &   1.3  &  -0.40  &  11.32  &   7.49  &   1.52  &   0.22 \\
    5 &    798 &   -20.85 &   5.9  &  -0.22  &  11.43  &   8.35  &   1.14  &   0.28 \\
    6 &    763 &   -21.69 &   1.5  &  -0.34  &  11.72  &   8.78  &   1.07  &  -0.03 \\
    7 &    731 &   -20.90 &   5.3  &  -0.20  &  11.45  &   8.72  &   0.97  &   0.50\\
    8 &   1535 &    ----- &   4.8  &  -0.06  &  -----  &   ----  &  -0.09  &   ----\\
    9 &   1903 &   -20.43 &   2.8  &  -0.21  &  11.27  &   8.60  &   0.93  &   0.20\\
   10 &   1632 &   -20.19 &   1.2  &  -0.11  &  11.20  &   8.73  &   0.84  &   0.46\\
   11 &   1231 &   -20.14 &   1.1  &  -0.27  &  11.13  &   10.06  &   0.13  &   0.25\\
   12 &   2095 &   -19.14 &   5.5  &  -0.41  &  10.69  &   9.93  &  -0.02  &   0.41\\
   13 &   1154 &   -19.96 &   1.6  &  -0.39  &  11.02  &   9.51  &   0.35  &   0.23\\
   14 &   1062 &   -19.79 &   2.7  &  -0.12  &  11.03  &   9.97  &   0.13  &   0.15\\
   15 &   2092 &   -19.72 &   5.4  &  -0.31  &  10.95  &   9.58  &   0.29  &  -0.05\\
   16 &    369 &   -18.63 &   2.7  &  -0.23  &  10.54  &   10.13  &  -0.19  &   0.69\\
   17 &    759 &   -19.51 &   1.6  &  -0.29  &  10.86  &   9.42  &   0.32  &   0.34\\
   18 &   1692 &   -19.20 &   5.4  &  -0.19  &  10.76  &   10.2  &  -0.12  &   0.31\\
   19 &   1030 &    ----- &   1.0  &  -0.22  &  -----  &   ----  &   0.90  &   ---- \\
   20 &   2000 &   -19.14 &   3.6  &  -0.54  &  10.64  &   10.00  &  -0.08  &   0.62 \\
   21 &    685 &   -19.26 &   4.6  &  -0.21  &  10.78  &   10.02  &  -0.02  &   0.41 \\
   22 &   1664 &   -19.22 &   1.7  &  -0.32  &  10.75  &   9.75  &   0.10  &   0.36 \\
   23 &    654 &   -18.87 &   4.7  &  -0.26  &  10.61  &   9.33  &   0.24  &  -0.01 \\
   24 &    944 &   -19.04 &   3.0  &  -0.27  &  10.68  &   9.87  &   0.00  &   0.22\\
   25 &   1938 &   -19.08 &   3.1  &  -0.65  &  10.60  &   9.65  &   0.08  &   0.36\\
   26 &   1279 &   -18.89 &   0.1  &  -0.47  &  10.56  &   9.83  &  -0.04  &   0.55\\
   28 &    355 &   -18.77 &   3.7  &  -0.18  &  10.60  &   10.01  &  -0.10  &   0.10\\
   29 &   1619 &   -18.67 &   1.2  &  -0.39  &  10.51  &   9.87  &  -0.08  &   0.23\\
   30 &   1883 &   -19.09 &   5.7  &  -0.49  &  10.64  &   9.23  &   0.30  &   0.24\\
   31 &   1242 &   -18.66 &   1.7  &  -0.40  &  10.49  &   9.43  &   0.13  &   0.55\\
   34 &    778 &   -18.18 &   2.7  &  -0.40  &  10.30  &   10.15  &  -0.32  &   0.48\\
   35 &   1321 &   -18.57 &   4.4  &  -0.40  &  10.46  &   8.71  &   0.47  &   0.19\\
   36 &    828 &   -18.14 &   1.3  &  -0.52  &  10.26  &   9.58  &  -0.06  &   0.59\\
   37 &   1250 &   -18.45 &   0.2  &  -1.03  &  10.25  &   9.55  &  -0.05  &   0.40\\
   38 &   1630 &   -18.31 &   1.2  &  -0.19  &  10.42  &   9.57  &   0.03  &   0.12\\
   40 &   1025 &   -18.24 &   4.3  &  -0.53  &  10.28  &   9.58  &  -0.05  &   0.72\\
   41 &   1303 &   -18.17 &   3.4  &  -0.35  &  10.30  &   9.33  &   0.09  &   0.44\\
   42 &   1913 &   -17.95 &   5.5  &  -0.10  &  10.29  &   9.35  &   0.07  &   0.45\\
   44 &   1125 &   -18.00 &   0.8  &  -0.65  &  10.15  &   9.30  &   0.03  &   0.54\\
   45 &   1475 &   -17.79 &   3.9  &  -0.76  &  10.05  &   9.48  &  -0.12  &   0.83\\
   46 &   1178 &   -17.79 &   4.2  &  -0.39  &  10.15  &   9.90  &  -0.28  &   0.75\\
   47 &   1283 &   -17.83 &   1.2  &  -0.44  &  10.15  &   8.97  &   0.19  &   0.59\\
   48 &   1261 &   -17.56 &   1.6  &  -1.01  &   9.91  &   8.66  &   0.23  &   0.69\\
   49 &    698 &   -17.81 &   2.0  &  -0.84  &  10.04  &   9.02  &   0.11  &   1.00\\
   50 &   1422 &   -17.25 &   2.2  &  -0.95  &   9.80  &   8.62  &   0.19  &   0.60\\
   53 &      9 &   -17.04 &   5.5  &  -1.34  &   9.65  &   8.09  &   0.38  &   0.75\\
   57 &    856 &   -16.60 &   2.6  &  -1.22  &   9.49  &   8.54  &   0.07  &   1.05\\
   60 &   1087 &   -16.98 &   0.9  &  -1.34  &   9.62  &   8.46  &   0.18  &   1.03\\
   62 &   1861 &   -16.63 &   2.8  &  -0.98  &   9.55  &   8.58  &   0.09  &   1.00\\
   63 &   1431 &   -16.73 &   3.1  &  -1.08  &   9.57  &   8.49  &   0.14  &   1.25\\
   63 &    543 &   -16.73 &   3.1  &  -1.49  &   9.50  &   8.42  &   0.14  &   0.80\\
   67 &   1833 &   -16.39 &   4.2  &  -0.95  &   9.46  &   9.11  &  -0.23  &   0.84\\
   68 &    437 &   -17.03 &   5.6  &  -1.72  &   9.60  &   8.19  &   0.30  &   0.80\\
   71 &    200 &   -16.06 &   3.5  &  -1.72  &   9.21  &   8.39  &   0.01  &   0.96\\
   73 &     21 &   -16.06 &   5.5  &  -1.72  &   9.21  &   8.62  &  -0.10  &   0.96\\
   78 &   1545 &   -16.08 &   0.9  &  -0.80  &   9.36  &   8.65  &  -0.04  &   1.34\\
   81 &   1075 &   -15.89 &   2.2  &  -1.60  &   9.15  &   8.22  &   0.07  &   1.15\\
   84 &   1440 &   -16.18 &   3.1  &  -0.27  &   9.53  &   9.14  &  -0.20  &   0.95\\
   85 &    230 &   -15.40 &   3.3  &  -1.72  &   8.95  &   8.36  &  -0.11  &   1.35\\
   89 &   1828 &   -15.76 &   2.3  &  -1.49  &   9.11  &   8.18  &   0.07  &   1.19\\
   91 &   1407 &   -15.82 &   0.6  &  -0.95  &   9.23  &   8.61  &  -0.09  &   1.42\\
   95 &   1539 &   -15.70 &   0.9  &  -1.16  &   9.14  &   7.69  &   0.33  &   1.41\\
   96 &   1185 &   -15.61 &   0.4  &  -1.72  &   9.03  &   7.89  &   0.17  &   1.37\\
\hline
\end{tabular}
\label{parametros}
\end{table*}

\section{Chemical abundances}
\label{CHAB}
\begin{figure*}
\resizebox{0.4\hsize}{!}{\includegraphics{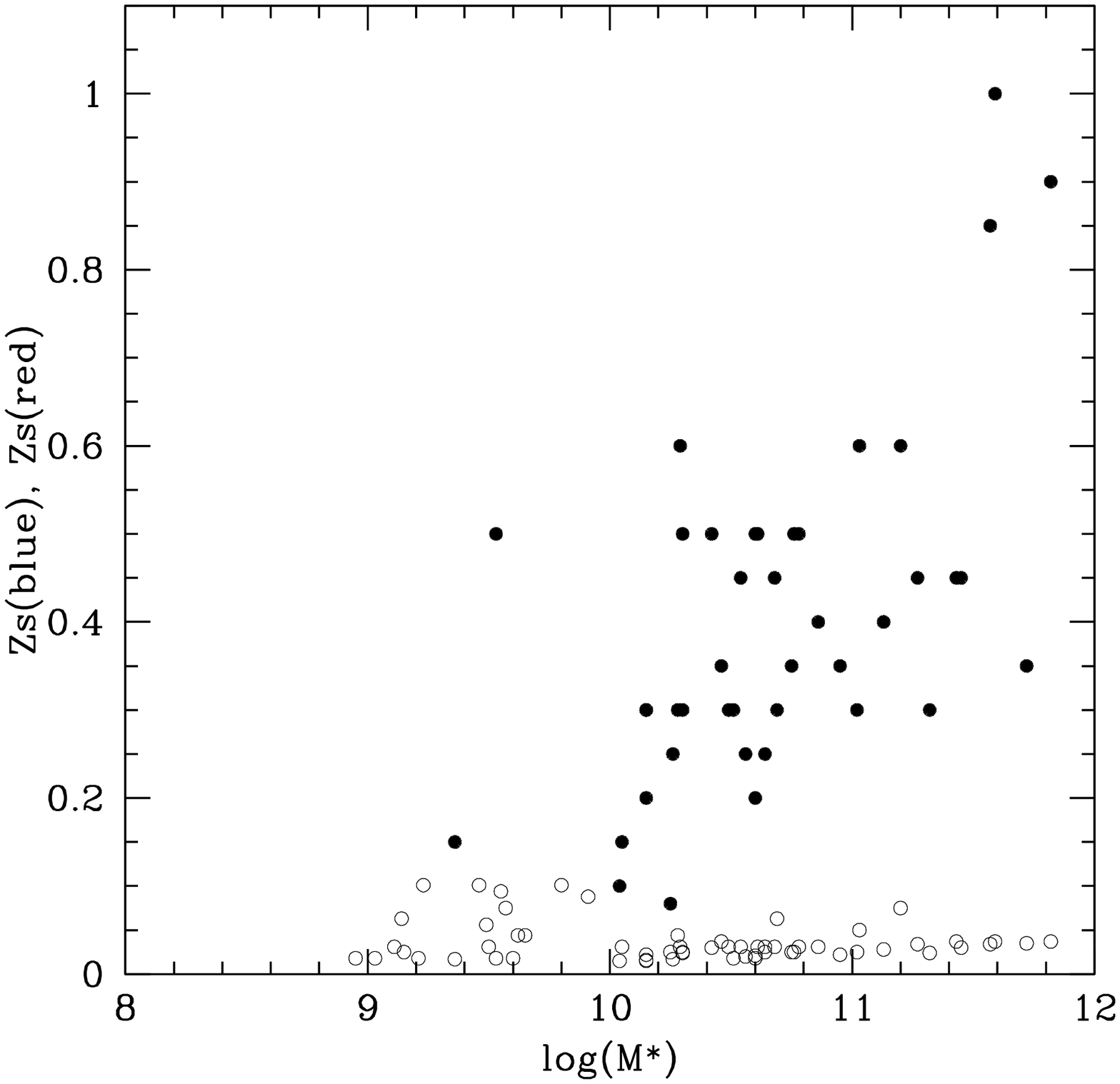}}
\resizebox{0.4\hsize}{!}{\includegraphics{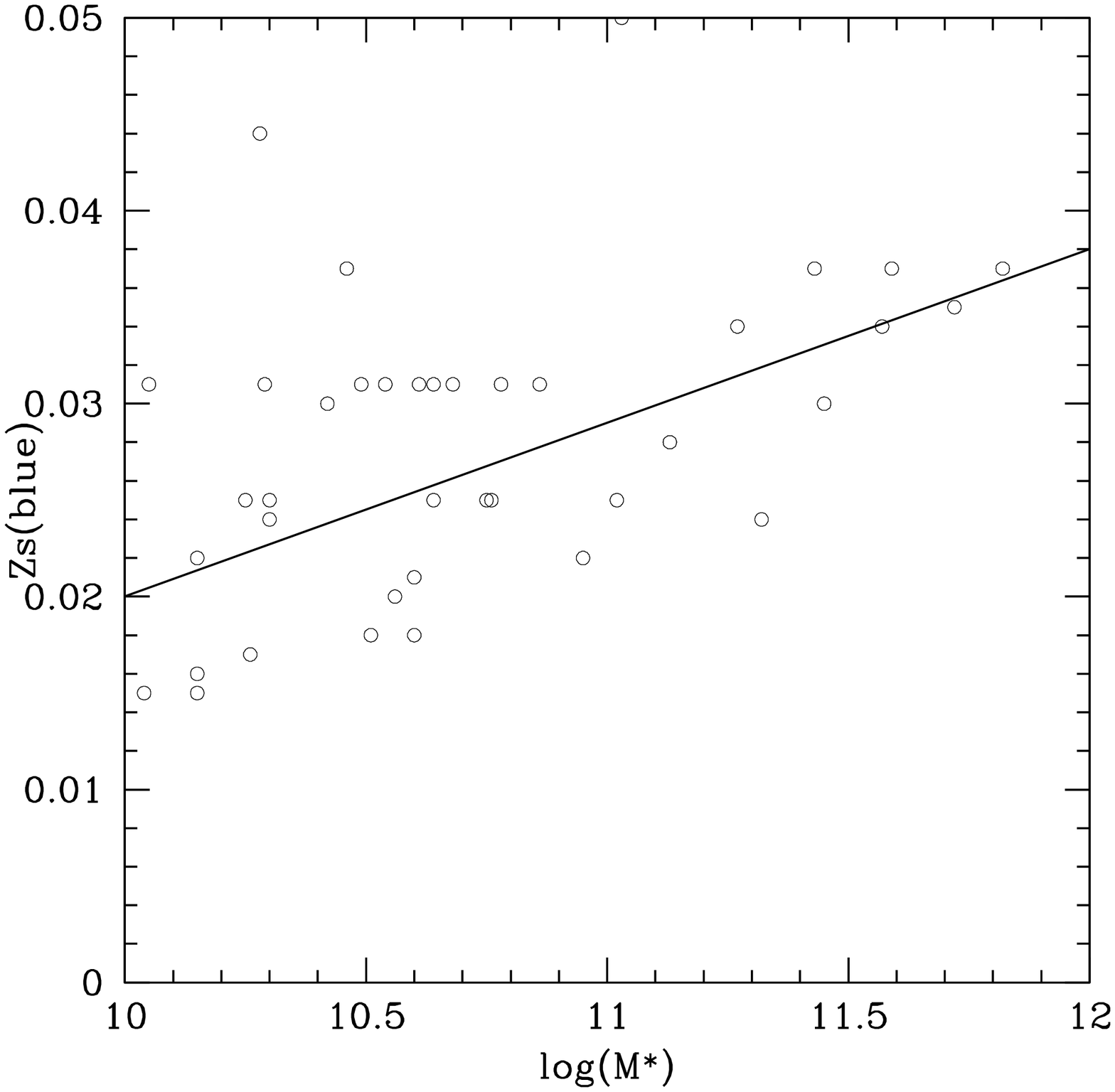}}
\caption{Left panel:
 Chemical abundance scales $Zs(blue)$ and $Zs(red)$ for the blue and red GC (open
 and filled dots, respectively) as a function of  galaxy stellar mass. Blue GC
 show a mild increase with mass consistent with previous results (e.g. Strader
\& Brodie 2007). Right panel: $Zs(blue)$ values, for galaxies more massive than 
$M_* \approx 10^{10} M_{\odot}$, as a function of galaxy stellar mass on an expanded scale.
 The straight line is a least square fit to the data points.
  }
\label{zsvsmass}
\end{figure*}

     The $Zs(blue)$ and $Zs(red)$ parameters (from Table\,\ref{sample}) are depicted
     as a function of total stellar mass in  Figure \ref{zsvsmass} (left panel). This
     diagram shows that the chemical abundance scale of the blue GC keeps
     mostly below $0.05~Z_{\odot}$. Five galaxies with $M_*$ smaller
     than $10^{10}~M_{\odot}$ (numbers 50, 91, 62, 67, 95), among the 18 galaxies
     that exhibit unimodal histograms, show rather high $Zs(blue)$ values. Their
     position in that diagram suggests they might be in fact bimodal, i.e, include
     a red GC component not much different in chemical abundance scale from that of 
     the blue GC and then broadening the histogram.

     For stellar masses larger than $10^{10}~M_{\odot}$ there seems to be a mild
     increasing trend of $Zs(blue)$ with galaxy mass as shown in Figure \ref{zsvsmass},
     right panel. This diagram also shows a linear least squares fit:
     $ log(Zs(blue))=-0.07 + 9.10^{-3}~log(M_*)$.
     This fit does not include the five galaxies mentioned before as is restricted to
     galaxies with stellar masses larger than $10^{10}~M_{\odot}$.
   
     This trend is coherent with \citet{SBF04} (see their figure 1)
     who find an increase of the mean $(V-I)$ colours of blue GC with increasing
     galaxy brightness. 

     Figure \ref{zsvsmass} also
     shows that the GC $Zs(red)$ parameters (filled dots) clearly correlate with galaxy
     mass. This result supports the idea that red GC can reach higher metallicities in
     the deeper potential well of the more massive systems. The relatively large
     spread of this relation, in turn, may be reflecting different degrees of 
     heterogeneity connected with the formation of the red GC and their associated
     diffuse stellar population. 

\begin{figure*}
\resizebox{0.4\hsize}{!}{\includegraphics{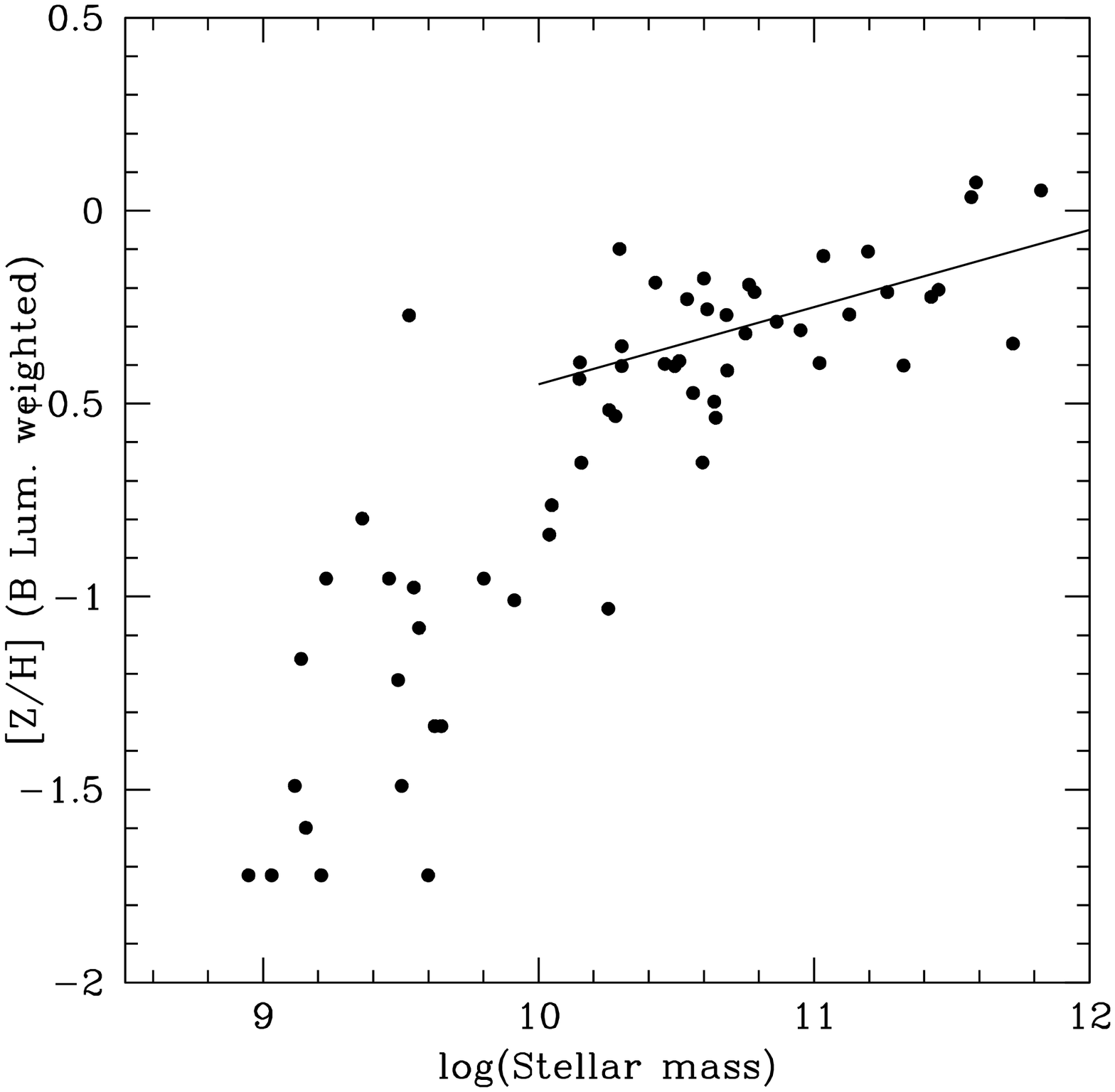}}
\resizebox{0.4\hsize}{!}{\includegraphics{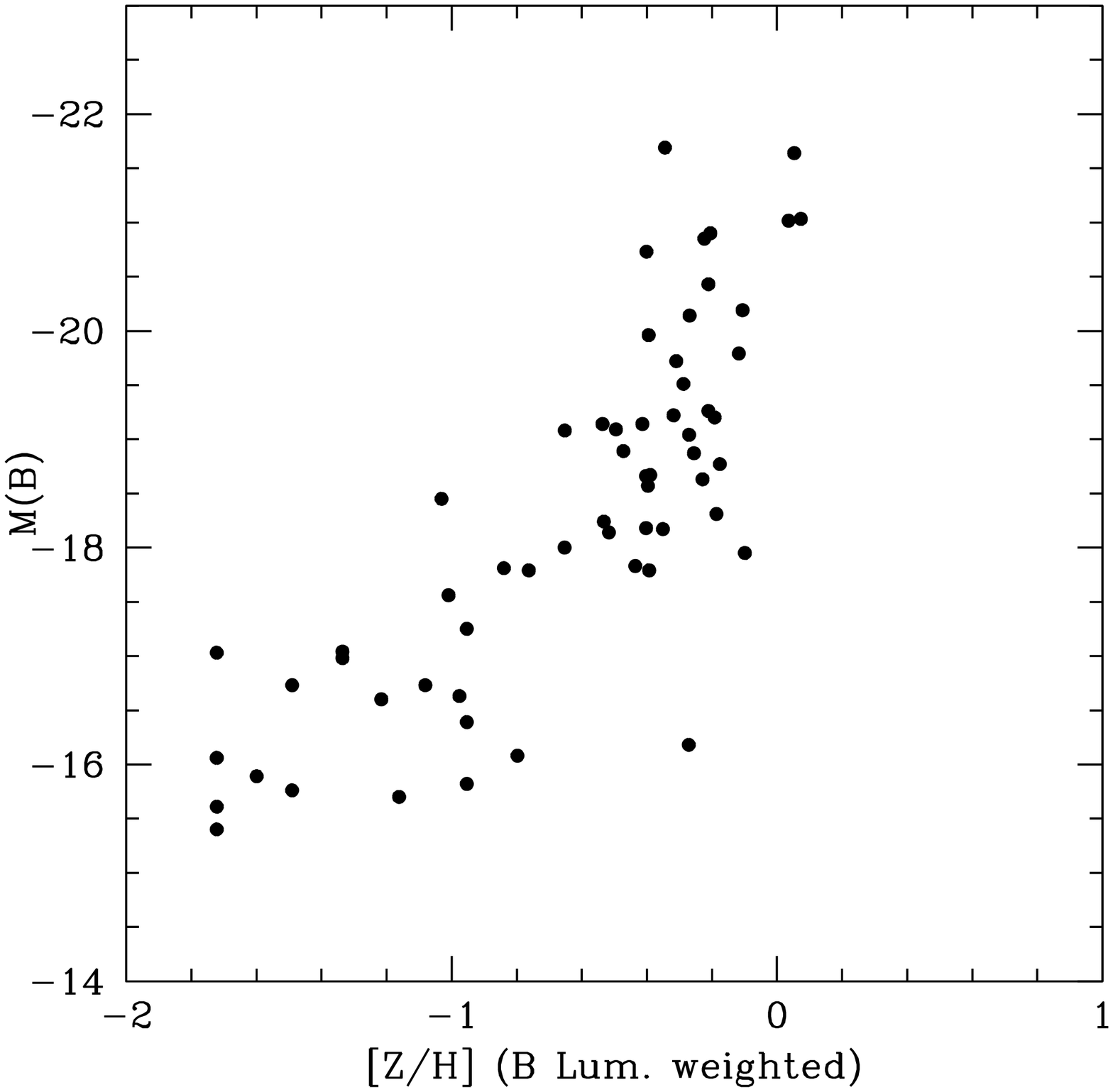}}
\caption{Left panel:
 Blue luminosity weighted chemical abundance as a function of total
   stellar mass. The straight line as a slope of -0.2, comparable to values
   reported in the literature. A significant change in that slope appears
   for masses smaller than  $M_* \approx 10^{10} M_{\odot}$. Right panel:
   Absolute $B$ magnitudes (within {\bf a}= 150 arcsecs)
   versus chemical abundance (B luminosity weighted).}
\label{zhmass}
\end{figure*}

     Blue luminosity weighted abundances $[Z/H]$, derived from the model
     are shown as a function of stellar mass and absolute
     $B$ magnitude in Figure \ref{zhmass}, and listed in Table\,\ref{parametros} (column 5).
     The first of these figures also includes a straight line with
     a slope $d[Z/H]/dlog(M_*) = -0.20$  , comparable to values found
     in the literature (see \citealt{LBA08},
     \citealt{Kod98}, \citealt{Tho05}), and gives a good approximation
     covering the upper two dex in stellar mass. However, both diagrams show
     a change in the slopes at a stellar mass $M_*\approx 10^{10} M_{\odot}$ or
     $M(B)\approx -17.5$. 
     We note below that stellar mass, the GC colour distribution becomes
     predominantly unimodal, i.e., the red GC component seems very
     small or is directly absent. The referred mass then seems then a key
     parameter regarding the type of stellar population present in
     a given galaxy. 


\section{The GC formation efficiency-galaxy mass-projected mass density space}
\label{GCFE}

      The  GC formation efficiency {\bf t} can be defined as a function
      of the total numer of GC and total stellar mass $M_*$ or, discriminating both
      GC families, as ${\bf t}_{red}=N_{red}/M_*$ and ${\bf t}_{blue}=N_{blue}/M_*$.
      The behavior of these three parameters as a function of the galaxy structural
      parameters is analyzed in what follows.

\begin{figure*}
\resizebox{0.4\hsize}{!}{\includegraphics{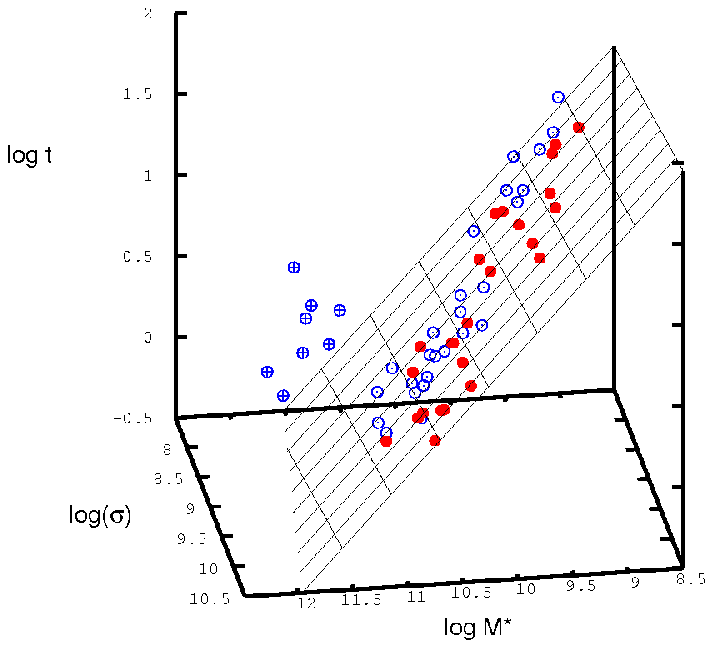}}
\resizebox{0.4\hsize}{!}{\includegraphics{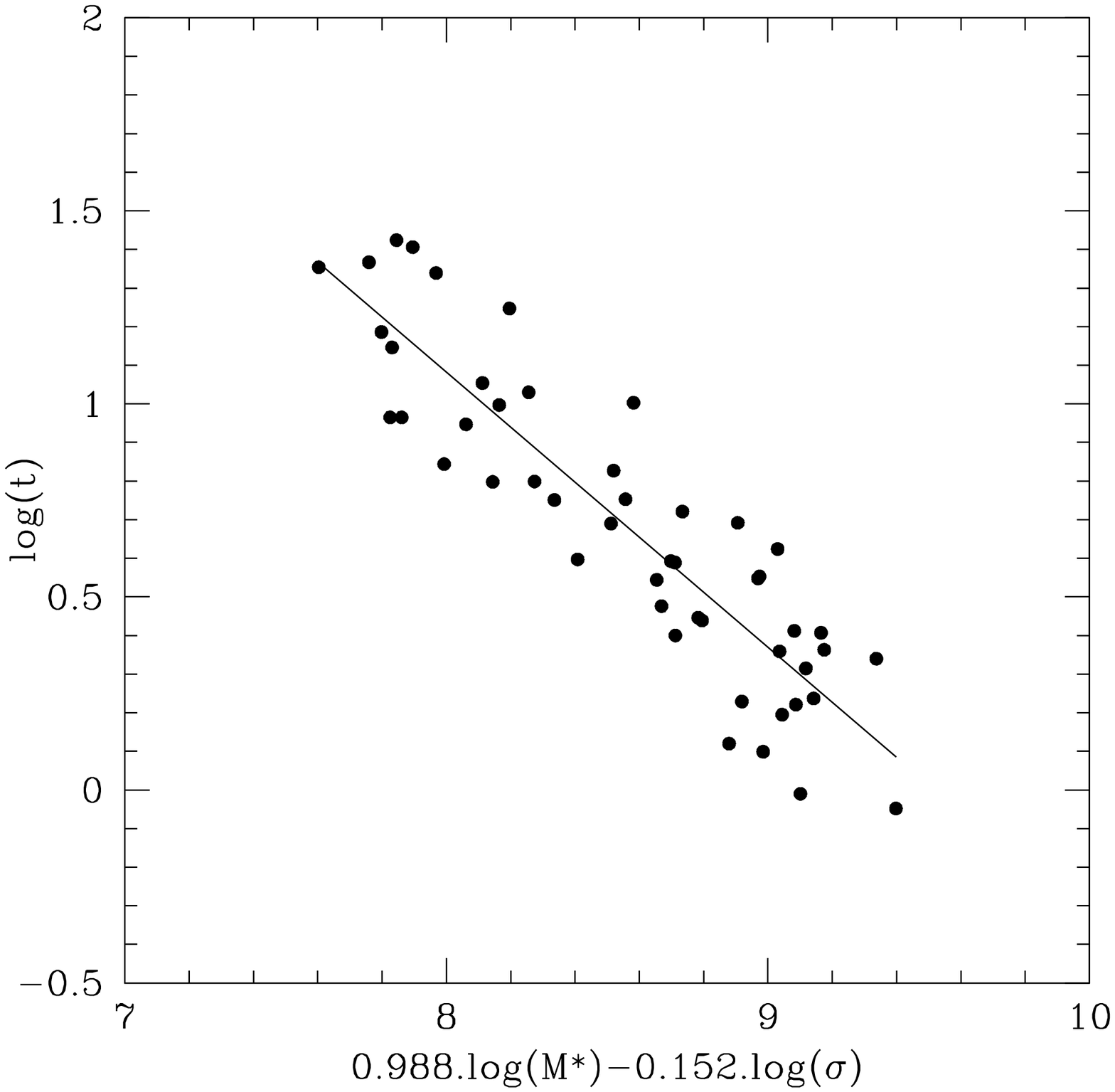}}
\caption{Left panel:
 Globular cluster efficiency (total number of clusters per stellar mass 
 unit) as a function of stellar mass and projected stellar mass
 density ($\sigma$). Filled dots lie below the plane. Dots with a 
 cross belong to the most massive galaxies. Right panel: Edge on view of 
 the plane depicted
 in the previous figure, for galaxies with total stellar mass smaller than
 $10^{11}~M_\odot$.}
\label{plane}
\end{figure*}

      Figure \ref{plane} (left panel) shows the (logarithmic) volume space defined by
      stellar mass, projected stellar mass density $\sigma$ (within the effective 
      radius $r_{e}$ and in solar masses per sq. Kpc), and total GC efficiency, {\bf t} 
      (all values listed in Table\,\ref{parametros}). These parameters were derived by 
      adopting our stellar mass estimates, the 
      galaxy effective radii from F2006, and the number of GC listed in Table\,\ref{sample}.


      This figure shows that most galaxies with stellar masses below
      $\approx 10^{11}$ $M_{\odot}$  define a thick plane. In turn, eight of the
      brightest galaxies appear clearly
      detached and above that plane. These last galaxies exhibit
      intermediate to low projected mass densities and, also, low central
      surface brightness compared to the S\'ersic profile that fits the outer
      galaxy profile (see \citealt {Kor08}). We point out that the {\bf t} parameters
      listed in Table\,\ref{parametros} for these galaxies are only approximate values. This is due
      to the fact that both GC families have cores in their spatial distribution (i.e.
      the projected GC density profiles are shallower than the galaxy brigtness profiles) leading to low
       {\bf t} ratios in their inner regions. On the other side, 
       as noted before, their total masses include a considerable fraction 
       beyond {\bf a}=150 arcsecs. A tentative correction for this last effect 
       would increase their tabulated $log({\sigma})$ values by $\approx 0.25$.

      A least square fit to the {\bf t} values (in $10^{-9}$ units) and assuming no
      errors on effective radii or projected mass density, yields, for 49
      galaxies with stellar masses smaller than $10^{11} M_{\odot}$:

\begin{eqnarray}
log(t)=-0.70(\pm0.05)~log(M_*)+0.11(\pm 0.04)~log(\sigma)+ \nonumber \\
6.75(\pm0.04)\               
\end{eqnarray}

\noindent with a fit rms of $\pm 0.17$ in $log(t)$. The fit residuals in log(t) 
show no correlation with galaxy mass, integrated colour, morphology or position within the Virgo cluster.

      An ``edge-on'' view of the plane is depicted in Figure \ref{plane} (right panel)
      excluding the eight brightest galaxies in Table\,\ref{sample}.

      The dependency with projected mass density becomes more evident if
      only the red GCs (in 31 bimodal galaxies), are included in the fit:

\begin{eqnarray}
log(t_{red})=-0.65(\pm 0.16)~log(M_*)+0.21(\pm 0.09)~log(\sigma)+\nonumber \\
4.99(\pm0.05)\              
\end{eqnarray}
\noindent with a fit rms of $\pm 0.21$ in $log(t_{red})$.

      On the other side, blue GC in 49 galaxies (31 with bimodal and 18 with unimodal 
      GC systems) show a steeper slope with stellar mass but no detectable dependence with
      projected mass density:

\begin{eqnarray}
log(t_{blue})=-0.91(\pm0.07)~log(M_*)+0.02(\pm0.05)~log(\sigma)+\nonumber \\
9.37(\pm0.04)\            
\end{eqnarray}

\noindent with a fit rms  of $\pm$ 0.19 in $log(t_{blue})$.

       This lack of dependency may be expected as the dominant stellar mass 
       in bimodal GC galaxies appears associated with the red GC population
       according to Table\,\ref{sample}.

\section{Conclusions}
\label{IPF}

   The analysis presented in this work suggests that GC in
   early type galaxies are good tracers of their stellar
   populations. We stress that, the proposed GC formation efficiency
   dependence with chemical abundance, leads to a good approximation
   of the galaxy integrated colours and is consistent with a constant
    $\delta$ parameter  suggesting a common pattern for the GC-field stars 
   connection.

    We also note, in particular, that this approach is able to explain the
   stellar halo-mean GC colour differences without implying definitely
   distinct formation histories for field stars and GC, although see 
  \citealt{Peng08} for a different interpretation.

   In turn, the inferred galaxy $(M/L)_{B}$ ratios compare well with the range 
   of values found in the literature and seem consistent with old ages and 
   small formation time spreads. 

   The on-set of GC colour bimodality seems to occur at masses
   larger than $M_* \approx 10^{10} M_{\odot}$, i.e., galaxies with lower masses
   only give rise to blue GC. Interestingly, \citet{DW}
   suggest that galaxies with masses smaller than
   $\approx 3.10^{10} M_{\odot}$ are efficiently cleaned of interstellar
   matter as a result of stellar and SNe winds leading to an
   interruption of the star formation process.

   The two dominant GC families have rather distinct characteristics.
   On one side, ``blue'' globulars can be characterized by a
   very low chemical abundance spread suggesting that some
   kind of major common event may have played a role in their
   ``synchronous'' formation and relatively homogeneous chemical
   composition. The re-ionization of the Universe has been put
   forward as possible responsible for this characteristic (e.g.
   \citealt{SAN03}) although the impact of re-ionization on galaxy and
   galaxy-cluster scales should also deserve further analysis. In any case,
   the increase of $Zs(blue)$ with galaxy mass lends support to the idea that
   blue GC in fact ``know'' about the galaxy they are related to (e.g.
   \citealt{SBF04}).

   The very narrow metallicity scale of the blue GC is compatible
   with their formation in low effective-yield environments
   characterized by important mass loss leading to a rapid
   supression of field star formation (e.g. \citealt{HARR06})
   and to high GC formation efficiencies. 

   In turn, red clusters exhibit an abundance spread that becomes
   larger for more massive galaxies and appear associated, when GC
   bimodality  is observable, with the dominant stellar mass
   component. This is coherent with the idea that more massive systems
   exhibit higher mean chemical abundances. The large dispersion of the
   relation between the $Zs(red)$ scale and galaxy mass may be also
   indicating some diversity in the formation history of this component
   (e.g. different numbers of mergers and their effects on gas removal).

   The overall increase of chemical abundance with mass
   derived in this work, is
   consistent with the GC formation modelling presented by \citet{BEK08}. 
   These models imply a formation time spread of roughly 3 Gy
   (i.e. $\pm$ 1.5 Gy around a mean value of about 11 Gyr. 

   We also stress that the broad behaviour derived for {\bf t} with galaxy mass is coherent
   with the starting hypothesis in this work, i.e., that the number of clusters per diffuse 
   stellar mass unit increases with decreasing metallicity (and hence with 
   decreasing stellar mass according to the stellar mass-metallicity relation).
 
   In more detail, galaxies with stellar masses lower that $M_* \approx 10^{11} M_\odot$ 
   define a thick plane in the logarithmic stellar mass-$\sigma$-{\bf t} space 
   where the  GC formation efficiency seem dependent on both galaxy
   mass and projected surface mass density. 
   Such a dependence becomes more evident when only red GC  are taken into
   consideration. This result resembles those by \citet{LR00} who suggest a dependence
   of the formation of young and massive stellar cluster  with projected mass density. 
   The lack of such a dependence in the case of the blue GC reinforces the idea
   that these  clusters have genetic differences with the red ones. 

   Remarkably,  the most massive galaxies in the sample appear clearly
   detached and above that plane. Naively, this is the expected situation for
   galaxies suffering mergers, although a number of dynamical effects that may
   have an impact on the {\bf t} parameter still remain to be considered before
   adopting this conclusion. For example,
   in the case of (dry) merging two galaxies with similar masses and GC populations,
   the composite {\bf t} parameter would remain unchanged while the 
   total mass would be doubled. These massive galaxies also exhibit
   large effective radii and intermediate projected stellar mass densities,
   a signature of major merger events (e.g. \citealt{SHIN}).

   A tentative conclusion, that will deserve further analysis, suggests that
   galaxies with stellar masses below that value, and located on the ``thick'' plane
   described in this paper, may have suffered a relatively small number of disippationless 
   merging events along their life. This conclusion also gets support from
   CDM based modelling (e.g. \citealt{DL06}) showing that less
   massive galaxies may have had a low number of galaxy-progenitors
   in their formation history.

\section*{Acknowledgments}
     This work was supported by grants from La Plata National University,
     Agencia Nacional de Promoci\'on Cient\'ifica y Tecnol\'ogica, and CONICET,
     Argentina. JCF acknowledges Dr. Michael West hospitality at ESO during the
     last stages of this paper. Useful comments and advise from an anonymous
     Referee are also acknowledged.

\section*{Appendix:GC colour histograms}  

GC colour histogram fits for 63 galaxies (*) in the Virgo ACS. Heavy black lines are the
(background corrected) observed histograms. Red lines show the histograms that give the best
fits in terms of the quality indicator (see Table\,\ref{sample}). Dots show the counting 
uncertainties for each colour bin. Note that Figure 4, in this paper, only shows nine representative
histogram fits.

\noindent (*) In the print version of the journal, the figures for only the first 12 galaxies are
included. The figures for all 63 galaxies can be found in the online version.

\begin{figure*}
\resizebox{1.0\hsize}{!}{\includegraphics{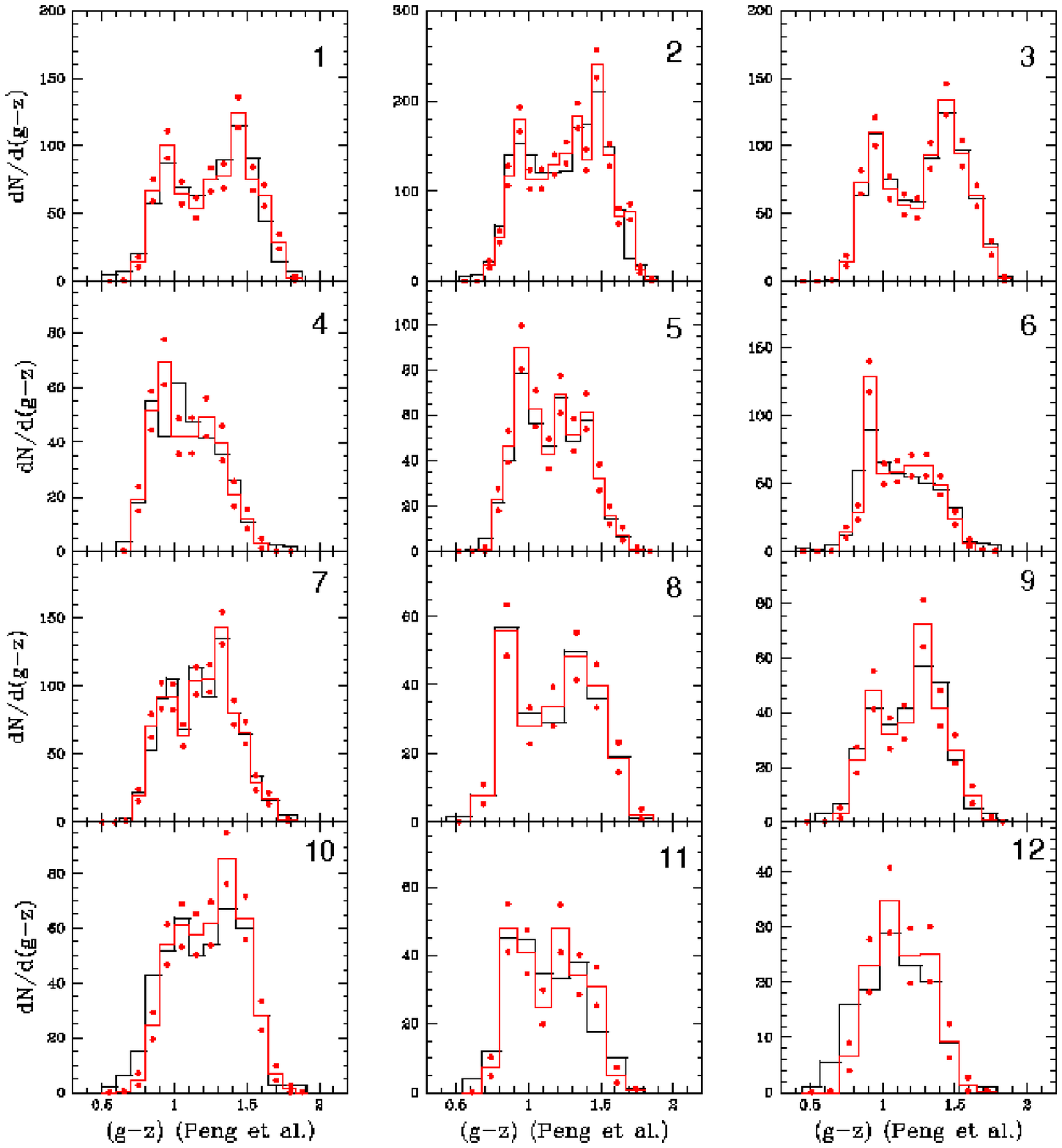}}
\end{figure*}

\begin{figure*}
\resizebox{1.0\hsize}{!}{\includegraphics{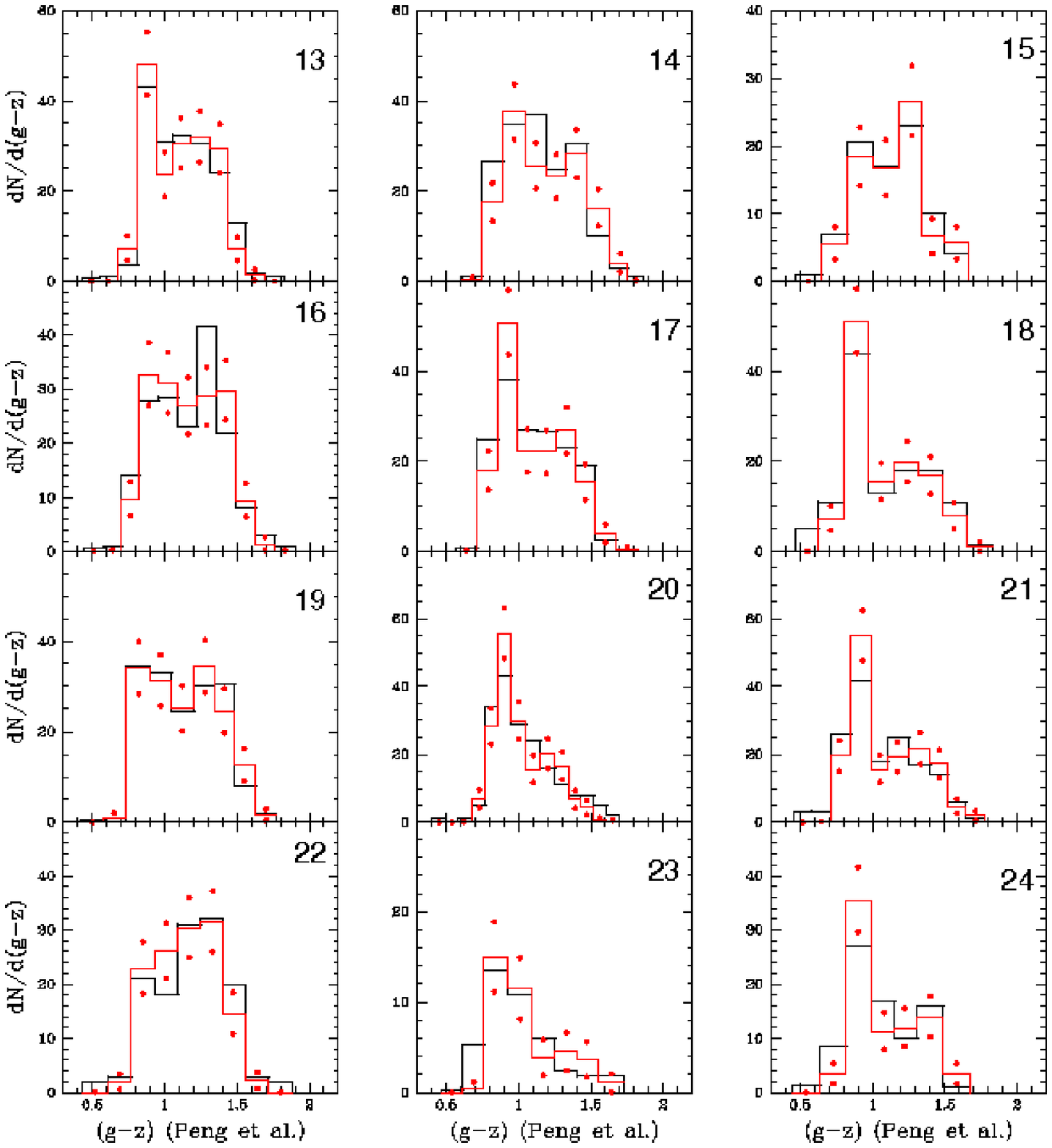}}
\end{figure*}

\begin{figure*}
\resizebox{1.0\hsize}{!}{\includegraphics{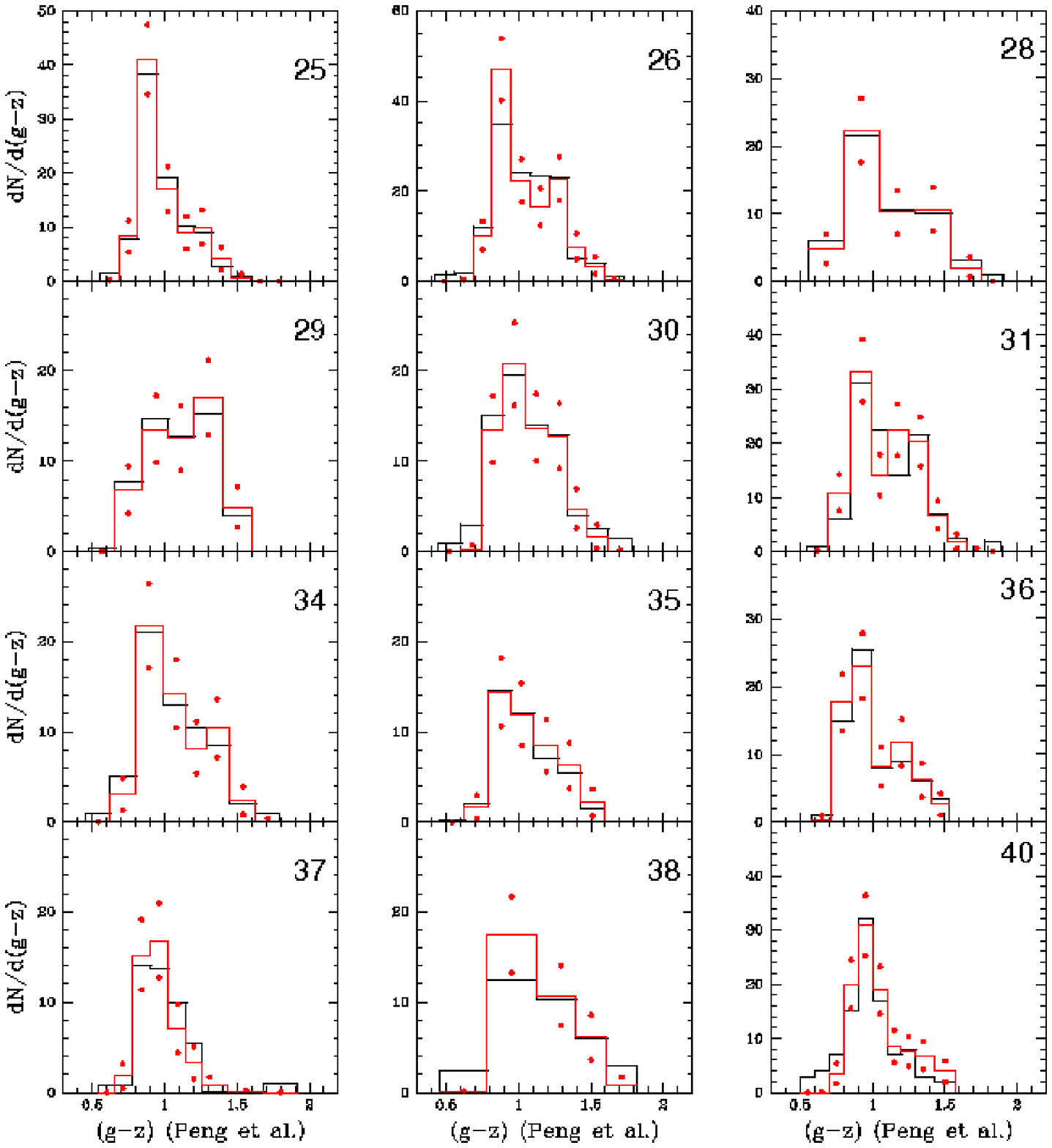}}
\end{figure*}

\begin{figure*}
\resizebox{1.0\hsize}{!}{\includegraphics{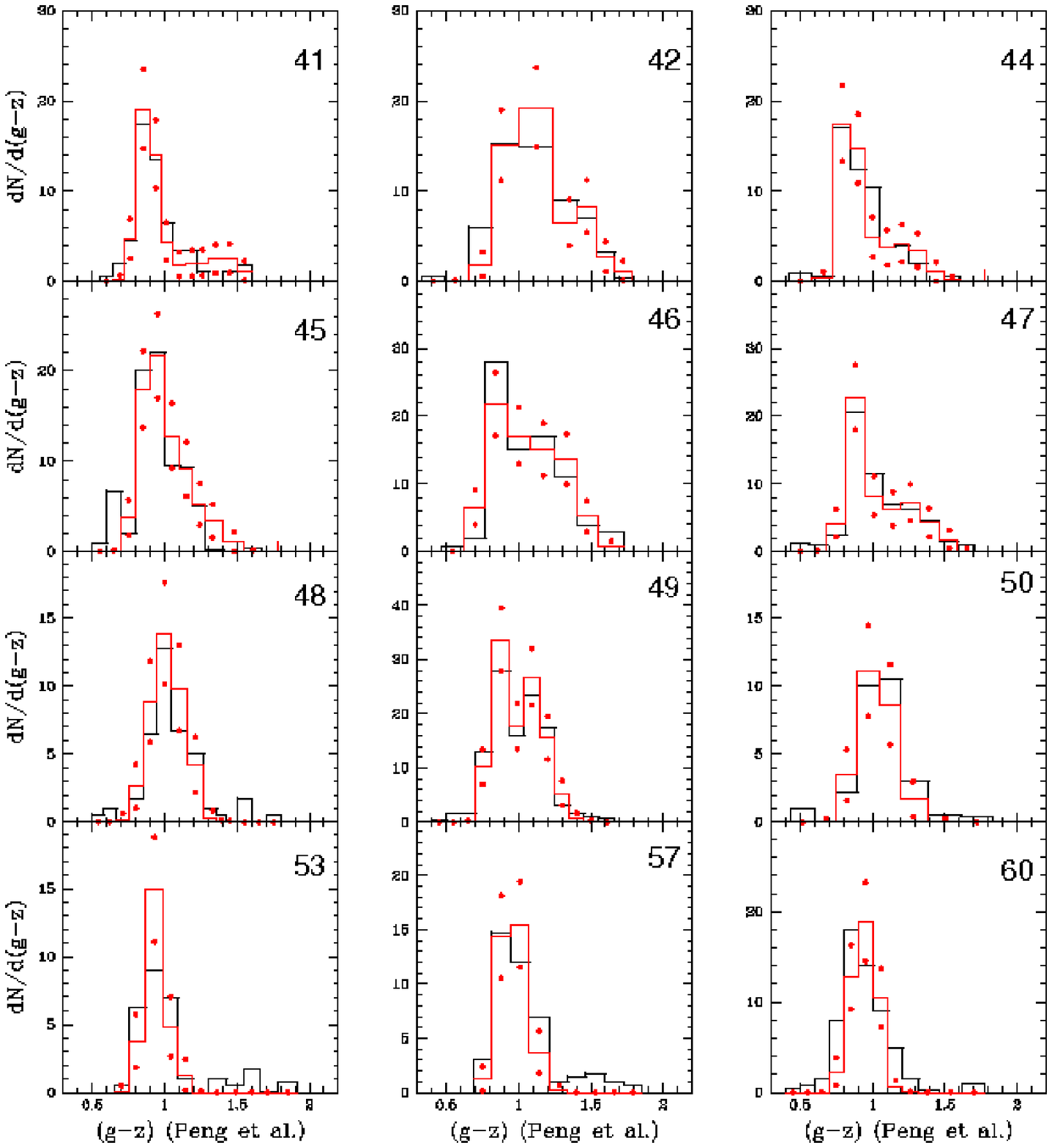}}
\end{figure*}

\begin{figure*}
\resizebox{1.0\hsize}{!}{\includegraphics{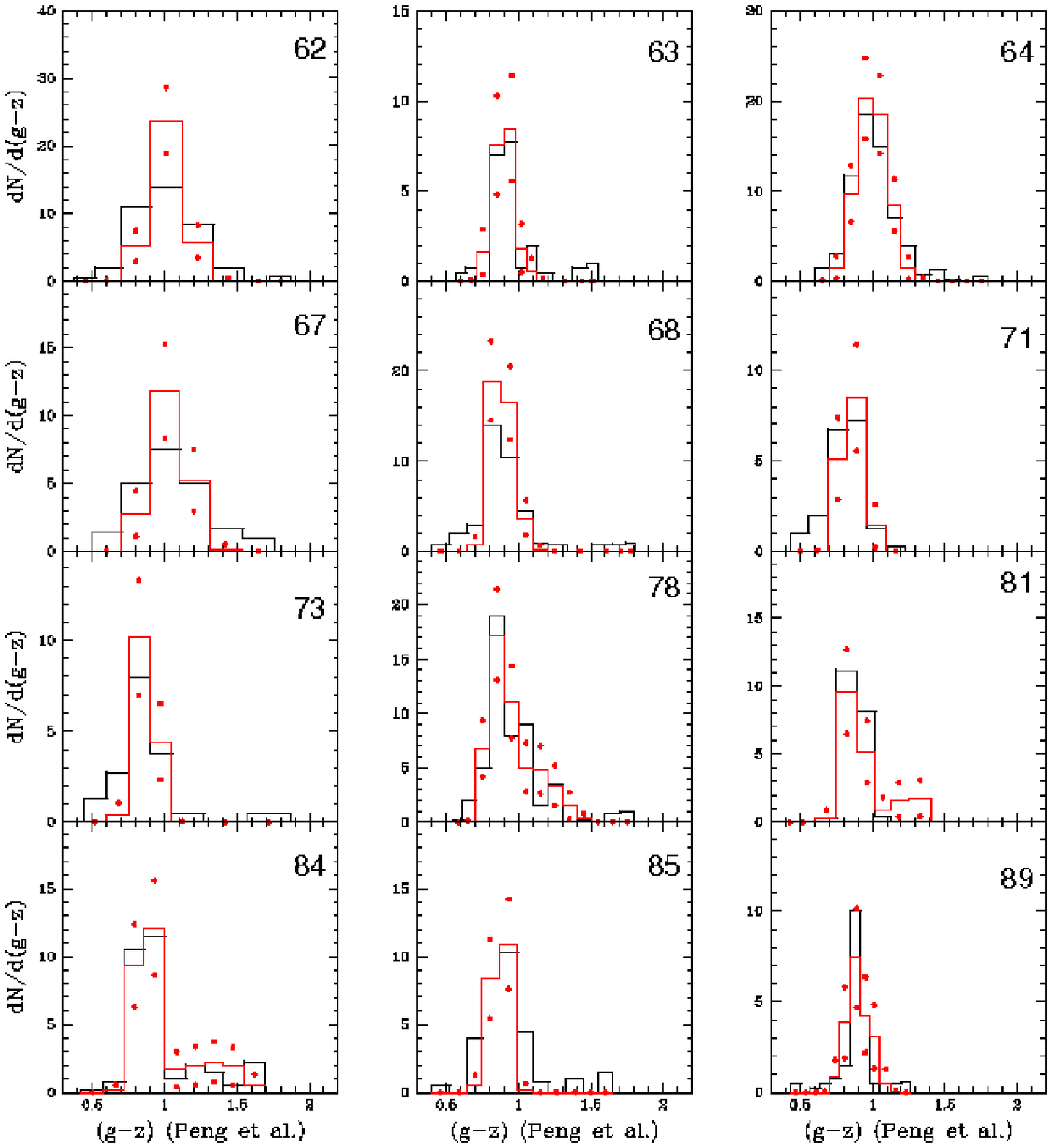}}
\end{figure*}

\begin{figure*}
\resizebox{1.0\hsize}{!}{\includegraphics{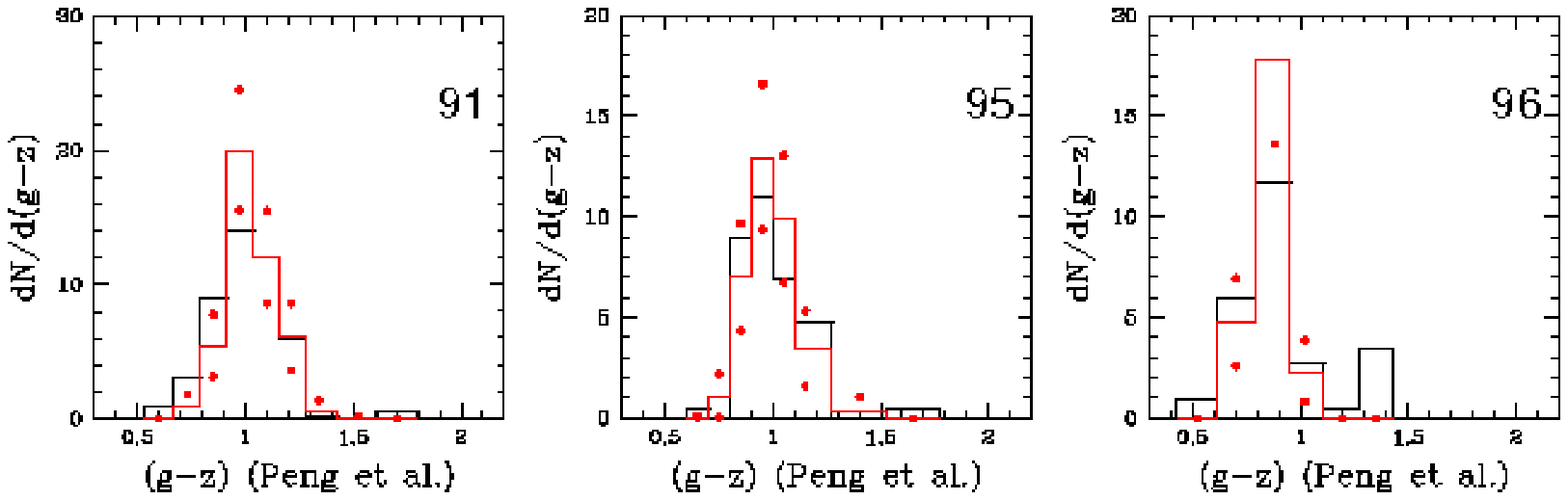}}
\label{lastpage}
\end{figure*}

\end{document}